\RequirePackage{fix-cm} % Allows arbitrary font scaling
\RequirePackage{silence} % Load silence BEFORE the document class
\documentclass[sn-mathphys,Numbered]{sn-jnl}% Math and Physical Sciences Reference Style

\usepackage{lmodern}

\usepackage{graphicx}%
\usepackage{multirow}%
\usepackage{amsmath,amssymb,amsfonts}%
\usepackage{amsthm}%

\usepackage[scr=rsfs]{mathalfa} 

\usepackage[title]{appendix}%
\usepackage{xcolor}%
\usepackage{textcomp}%
\usepackage{manyfoot}%
\usepackage{booktabs}%
\usepackage{algorithm}%
\usepackage{algorithmicx}%
\usepackage{algpseudocode}%
\usepackage{listings}%
\usepackage{bm}
\usepackage{mathtools}

\usepackage{caption}
\usepackage{subcaption}

\usepackage{hyperref}

\DeclarePairedDelimiter\abs{|}{|}
\newcommand{\sign}{\text{sign}}

\begin{document}

\title[Article Title]{Visualization and Analysis of the Curvature Invariants in the Alcubierre Warp-Drive Spacetime}

\author*[1]{\fnm{Jos\'{e}} \sur{Rodal}}\email{jrodal@alum.mit.edu}

\affil*[1]{\orgname{Rodal Consulting}, \orgaddress{\street{205 Firetree Ln.}, \city{Cary}, \state{NC}, \postcode{27513}, \country{USA}}}

\abstract{In the Alcubierre warp-drive spacetime, we investigate the following scalar curvature invariants: the scalar \textit{I}, derived from a quadratic contraction of the Weyl tensor, the trace \textit{R} of the Ricci tensor, and the quadratic \textit{r1} and cubic \textit{r2} invariants from the trace-adjusted Ricci tensor. In four-dimensional spacetime the trace-adjusted Einstein and Ricci tensors are identical, and their unadjusted traces are oppositely signed yet equal in absolute value. This allows us to express these Ricci invariants using Einstein’s curvature tensor, facilitating a direct interpretation of the energy-momentum tensor.  We present detailed plots illustrating the distribution of these invariants. Our findings underscore the requirement for four distinct layers of an anisotropic stress-energy tensor to create the warp bubble. Additionally, we delve into the Kretschmann quadratic invariant decomposition. We provide a critical analysis of the work by Mattingly et al., particularly their underrepresentation of curvature invariants in their plots by 8 to 16 orders of magnitude. A comparison is made between the spacetime curvature of the Alcubierre warp-drive and that of a Schwarzschild black hole with a mass equivalent to the planet Saturn. The paper addresses potential misconceptions about the Alcubierre warp-drive due to inaccuracies in representing spacetime curvature changes and clarifies the classification of the Alcubierre spacetime, emphasizing its distinction from class \textit{B} warped product spacetimes.}

\keywords{Alcubierre warp-drive, Curvature Invariants, Ricci tensor, Weyl tensor, Einstein tensor, Kretschmann scalar.}

\maketitle

\section{Introduction}\label{sec1}

\makeatletter
\newcommand\incircbin
{%
  \mathpalette\@incircbin
}
\newcommand\@incircbin[2]
{%
  \mathbin%
  {%
    \ooalign{\hidewidth$#1#2$\hidewidth\crcr$#1\bigcirc$}%
  }%
}
\newcommand{\oland}{\incircbin{\land}}
\makeatother

Scalar invariants play a pivotal role in general relativity  
\cite{Carminati,Harvey,Zakhary,Cherubini,Santosuosso,MacCallum}, offering insights that do not depend on arbitrary choices of coordinates or their transformations. One of the intriguing applications of these invariants is the study of the Alcubierre warp bubble \cite{Alcubierre_1994,AlcubierreLobo}. The Alcubierre metric, introduced in 1994 by Miguel Alcubierre \cite{Alcubierre_1994}, describes a spacetime geometry wherein a theoretical ``warp bubble'' moves superluminally, effectively allowing for faster-than-light travel of a spaceship inside the warp bubble.  Alcubierre's original paper has garnered 322 citations (as of August 2023 \cite{dimensions}) and the time rate of citations has been steadily increasing, highlighting its impact and the subsequent research it has stimulated. The Alcubierre warp-drive, as it is often termed, has been the subject of numerous studies examining its energy requirements, stability, and potential signatures \cite{Hiscock_1997,GonzalezDiaz,HiscockClark,LoboVisser,SantiagoVisser,Olum,Pfenning}. Notably, the solution has prompted discussions on the nature of energy conditions regarding ``exotic'' negative matter in general relativity, given that the warp-drive's feasibility hinges on the existence of negative energy densities \cite{LoboVisser,SantiagoVisser,Olum,Pfenning}.

Beyond its theoretical implications in Einstein's theory of general relativity, the Alcubierre metric has also been explored in the context of advanced propulsion concepts. While the practical feasibility of such a drive remains speculative, the underlying mathematics offers insights into the possible malleability of spacetime under specific matter-energy distributions. This has led to further investigations into geometries that might permit effective superluminal motion, even if the energy conditions for the Alcubierre drive prove insurmountable \cite{Lentz_2021,Fell,VanDenBroeck}.  In this article, we undertake a detailed examination of the curvature invariants associated with the Alcubierre metric. By analyzing their amplitude and spatial distribution, we shed light on the intrinsic spacetime characteristics of the warp bubble. Our findings provide insights into the unique properties of the Alcubierre warp-drive and contextualize them with conclusions that are invariant under coordinate transformations. \par
Mattingly et al. \cite{Mattingly} are, to the best of our knowledge, the sole contributors to the literature on the curvature invariants of the Alcubierre warp-drive spacetime. Their research recognized the value of intrinsic characterization of the Alcubierre warp-drive spacetime using curvature invariants, emphasizing their independence from coordinate choice. However, their publication exhibits problems in the graphical portrayal of these invariants. Specifically, the omission of plots delineating the invariant distribution of curvature invariants for the Alcubierre metric in planform and three-dimensional views, and most significantly their arbitrary truncation of the curvature invariants by 8 to 16 orders of magnitude, are underscored as major lapses that motivate this current study.  The imperative nature of representing curvature magnitude with precision is accentuated by drawing a parallel with the spacetime curvature of a Schwarzschild black hole. In particular, the curvature at the event horizon of a Schwarzschild black hole, equivalent in mass to Saturn, mirrors the apex amplitude of the curvature invariant distortion of spacetime in the Alcubierre warp-drive. The paper also elucidates that the truncated values shown by Mattingly et al. \cite{Mattingly} appear to be due to the inherent behavior of the Wolfram \textit{Mathematica\textsuperscript{\textregistered}} software, which auto-adjusts the range of plots. Such a representation could engender misconceptions regarding the viability and ramifications of the Alcubierre warp-drive paradigm, by giving the impression that the actual invariant curvature is smaller than its true value by 8 to 16 orders of magnitude.

Moreover, this article elucidates that the Alcubierre warp-drive line element does not categorically align with the Class $B_{1}$ Warped Product Spacetime, countering the claim made (without proof) by Mattingly et al. \cite{Mattingly}. The Alcubierre metric, predicated on the 3+1 formalism, is distinct from both the globally hyperbolic spacetime and Class $B$ spacetime attributes. The Alcubierre metric does not adhere to the Class $B$ spacetime constructs due to its form function coupling term that amalgamates all spacetime coordinates.

In summation, while the contributions of Mattingly et al. \cite{Mattingly} to the Alcubierre warp-drive spacetime pioneered the examination of the Alcubierre warp-drive spacetime using curvature invariants, their article contains important inaccuracies and misconceptions that necessitate rectification for a lucid understanding of curvature invariants of the Alcubierre metric.

\section{Preliminaries: Tensor Definitions}\label{sec2}

Tensors will be expressed using both index notation, e.g., $R_{\alpha \beta}$, and coordinate-free geometric notation, e.g., $Ric$, for clarity and comprehensive representation. We introduce several trace-adjusted tensors rendered traceless. Such tensors are denoted by a hat, e.g., $\widehat{T}^{\alpha \beta}$, distinguishing them from their unadjusted counterparts e.g., $T^{\alpha \beta}$.\par 
A trace-adjusted ($tr(\widehat{Ric})=0$) Ricci curvature tensor is defined as follows, in terms of the 2nd-rank Ricci curvature tensor $Ric$, Ricci curvature scalar $R\equiv tr(Ric)=R_{\alpha}\,^{\alpha}$, metric tensor $g$ and spacetime manifold dimension $n=g_{\alpha \beta} \: g^{\alpha \beta}=\delta_{\alpha}^{\alpha}$:

\begin{equation}
\widehat{Ric}\equiv Ric -\frac{1}{n} \: R \:g.\label{HatRic}
\end{equation}

For example, in four-dimensional spacetime ($n=4$), $\widehat{R}_{\alpha \beta}=R_{\alpha \beta}-\frac{1}{4} R \: g_{\alpha \beta}$.\par

Similarly a trace-adjusted Einstein curvature tensor ($tr(\widehat{Eins})=0$) is defined as follows, in terms of the 2nd-rank Einstein curvature tensor $Eins\equiv Ric - \frac{1}{2} R \: g$, Einstein curvature scalar $G\equiv tr(Eins)=G_{\alpha}\,^{\alpha}$,  metric tensor $g$ and spacetime manifold dimension $n$:

\begin{equation}
\begin{aligned}
\widehat{Eins} &\equiv Eins -\frac{1}{n} \: G \:g \\
&= Ric -\frac{n-2}{2n} \: R \:g\\
&= \widehat{Ric} +\frac{4-n}{2n} \: R \:g.
\end{aligned}
\label{HatEins}
\end{equation}

In four-dimensional spacetime ($n=4$), the trace-adjusted Einstein curvature tensor $\widehat{Eins}$ happens to be exactly equal to the trace-adjusted Ricci curvature tensor $\widehat{Ric}$, while the traces of the unadjusted tensors are equal in magnitude with opposite signs:

\begin{equation}
\begin{aligned}
\widehat{Eins} &= \widehat{Ric} \\
G &= -R
\end{aligned}
\quad \text{for } n = 4.
\label{EinsRiccN4}
\end{equation}

The (4th rank and invariant under conformal rescaling of the metric) traceless tensor $Weyl$ is defined as what remains of the (4th rank) Riemann curvature tensor $Riem$ after subtraction of the Kulkarni–Nomizu product \cite{andrews2010ricci} (denoted by $\oland$ ) of the (2nd rank) trace adjusted Ricci tensor $\widehat{Ric}$ with the (2nd rank) metric tensor $g$, and subtraction of the $\oland$ product of the metric tensor with itself:

\begin{equation}
\begin{aligned}
Weyl &\equiv Riem -\frac{1}{n-2} \widehat{Ric} \oland g-\frac{1}{2n(n-1)} g \oland g.\\
\label{weylDef}
\end{aligned}
\end{equation}

Eqs.~(\ref{weylDef}) and (\ref{EinsRiccN4}) can be combined to decompose the Riemann curvature tensor $Riem$ into the traceless tensor $Weyl$, the trace adjusted Einstein tensor $\widehat{Eins}$ and the metric tensor as follows:

\begin{equation}
\begin{aligned}
Riem &= Weyl +\frac{1}{2} \widehat{Eins} \oland g+\frac{1}{24} g \oland g
\quad \text{for } n = 4.
\label{RiemDec}
\end{aligned}
\end{equation}

We also define a trace-adjusted stress-energy-momentum tensor ($tr(\widehat{Tmom})=0$) as follows, in terms of the 2nd-rank stress-energy-momentum tensor $Tmom$, its contracted scalar $T\equiv tr(Tmom)=T_{\alpha}\,^{\alpha}$, and spacetime manifold dimension $n$:

\begin{equation}
\widehat{Tmom}\equiv Tmom -\frac{1}{n} \: T \:g.\label{HatT}
\end{equation}

For example, in four-dimensional spacetime ($n=4$), $\widehat{T}_{\alpha \beta}=T_{\alpha \beta}-\frac{1}{4} T \: g_{\alpha \beta}$.\par 

Einstein's field equations, when the cosmological constant is omitted due to its negligible effect on local scales (as opposed to cosmological scales), take the form $G_{\alpha \beta}=\kappa \: T_{\alpha \beta}$. In this form, the Einstein gravitational coupling parameter $\kappa$ serves as a spacetime compliance prefactor on the stress-energy tensor components $T_{\alpha \beta}$ to give the Einstein curvature tensor components $G_{\alpha \beta}$. Einstein’s field equations can be decomposed into an invariant scalar equation in terms of the traces $G$  and $T$, and a tensor equation in terms of traceless tensors (known in the continuum mechanics literature as ``deviatoric'' tensors) $\widehat{G}$  and $\widehat{T}$, with the inverse $\frac{1}\kappa{}$ of Einstein's gravitational coupling parameter acting as a spacetime stiffness:

\begin{equation}
\begin{aligned}
T &=\frac{1}{\kappa} G \\
\widehat{T}_{\alpha \beta} &= \frac{1}{\kappa}\widehat{G}_{\alpha \beta}
\end{aligned}
\quad \text{for } n = 4.
\label{FielEq}
\end{equation}\par

The single coupling parameter $\frac{1}{\kappa}$ in Einstein’s equation defines a direct, algebraic, homogeneous, and isotropic gravitational dependence between spacetime curvature $G_{\alpha \beta}$ and the stress-energy-momentum tensor $T_{\alpha \beta}$. For the Alcubierre warp-drive, the metric is predefined, allowing for the determination of its derivatives, and hence, predetermination of the curvature.  The scalar field equation, as given by the first equation in Eq.~(\ref{FielEq}), expressed in terms of the traces $T$ and $G$, provides the scalar gravitational source term $T$ responsible for curvature due to volume change (expansion or contraction) of spacetime.\par

The tensor field equation, represented by the second equation in Eq.~(\ref{FielEq}),  dictates the local shear stress, corresponding to the off-diagonal components of $\widehat{T}_{\alpha \beta}$  which in turn sources a shear curvature (warping) of spacetime (in the case of the Alcubierre metric this curvature is predefined). The stress-energy-momentum tensor can exhibit off-diagonal components in scenarios involving anisotropic fluids \cite{Herrera,AbellánVasilev,AbellánVasilevGRG,SantosCharged,SantosCosmol,SantosDust,SantosFluid,SantosGross}, which may arise due to fluid viscosity, or when observing a perfect fluid from a non-comoving frame, or in heat transfer. An electromagnetic field can also serve as a source in this context. Notably, the trace of the electromagnetic tensor is zero, implying that the electromagnetic field sources shear curvature of spacetime.

\section{Curvature Invariants in General Relativity}\label{sec3}

\subsection{Standard Curvature Invariants}\label{StdCurvatureInvariants}

Invariants play a crucial role in general relativity because their values remain unchanged under coordinate transformations \cite{Carminati,Harvey,Zakhary,Cherubini,Santosuosso,MacCallum}. They enable us to investigate the properties of spacetime, such as whether it is finite everywhere and under what conditions. For example, invariants can be used to determine whether a singularity is an essential physical phenomenon or merely a coordinate singularity that can be removed by an appropriate choice of coordinates. Curvature invariants are scalar quantities derived from curvature tensors, such as the Riemann, Weyl, and Ricci tensors. Operations like contraction, covariant differentiation, trace adjustment, and dualization are employed to form these invariants, using the metric tensor. Moreover, there are various methods to construct invariants. One common approach is to use polynomials, leading to invariants of linear, quadratic, cubic, quartic, quintic, and potentially higher orders, which are referred to as polynomial invariants. Although these polynomial invariants have been successfully used in many applications, they have limitations in distinguishing some Kundt spacetimes (Lorentzian manifolds admitting a geodesic null congruence where the following optical scalars: expansion, twist and shear all vanish). In particular, spacetimes with vanishing scalar invariants include certain Petrov type N (having one quadruple principal null direction, associated with the transverse gravitational waves detected by LIGO, and also with pp-waves) and Petrov type III spacetimes (having one triple and one simple principal null direction, associated with longitudinal gravitational radiation). These spacetimes cannot be differentiated from Minkowski spacetime using polynomial curvature invariants of any order.\par

One of the most important invariants of the Riemann tensor of a four-dimensional Einstenian manifold is the quadratic scalar invariant, formed from the contraction of the Riemann tensor with itself, known as the Kretschmann scalar, defined as:

\begin{equation}
\begin{aligned}
K &\equiv R_{\alpha\beta\gamma\delta} R^{\alpha\beta\gamma\delta}. \\
\end{aligned}
\label{Kretschmann}
\end{equation}\par

Where $R_{\alpha\beta\gamma\delta}$ and $R^{\alpha\beta\gamma\delta}$ are the fully covariant and the fully contravariant components, respectively, of the Riemann curvature tensor $Riem$.  The presence of a well-defined Riemann tensor implies that the Kretschmann scalar is also defined at the same location, and the converse holds true. Therefore, if the Kretschmann scalar can be evaluated at a specific point, it provides sufficient evidence that the geometry is regular at that coordinate location.\par

Another quadratic scalar invariant (which appears as a gravitational term in the Lagrangian in conformal gravity theories), formed from the contraction of the traceless Weyl curvature tensor with itself, is known as the Weyl scalar $I$ defined as:

\begin{equation}
\begin{aligned}
I &\equiv C_{\alpha\beta\gamma\delta} C^{\alpha\beta\gamma\delta}. \\
\end{aligned}
\label{WeylScalar}
\end{equation}\par

Where $C_{\alpha\beta\gamma\delta}$ and $C^{\alpha\beta\gamma\delta}$ are the fully covariant and the fully contravariant components, respectively, of the Weyl curvature tensor $Weyl$.\par

Similarly, we can define a quadratic scalar invariant (which also appears as a gravitational term in the Lagrangian in conformal gravity theories) in terms of the trace-adjusted Ricci curvature tensor (which for 4-dimensional spacetime can also be expressed in terms of the trace-adjusted Einstein curvature tensor using Eq.(\ref{EinsRiccN4})) as:

\begin{equation}
\begin{aligned}
r_{1} &\equiv \widehat{R}_{\alpha}^{\:\:\:{\beta}}\widehat{R}_{\beta}^{\:\:\:{\alpha}} \\
 &= \widehat{G}_{\alpha}^{\:\:\:{\beta}}\widehat{G}_{\beta}^{\:\:\:{\alpha}} \quad \text{for } n = 4.
\end{aligned}
\label{r1}
\end{equation}\par

This definition differs from Carminatti et al. \cite{Carminati} who define this quadratic traceless Ricci invariant  with an arbitrary prefactor of $1/4$ in the 4-dimensional Einstenian manifold.\par

The Kretschmann scalar is related to the other invariants by this expression (Cherubini et al.\cite{Cherubini}):

\begin{equation}
\begin{aligned}
K &= I+\frac{4}{n-2}{R}_{\alpha \beta} {R}^{\alpha \beta}-\frac{2}{(n-1)(n-2)} {R}^2\\
&= I+\frac{4}{n-2}\:{r}_{1} +\frac{4(n-1)-2n}{n(n-1)(n-2)} {R}^2.
\end{aligned}
\label{KIR}
\end{equation}\par

Therefore in the Einstenian four-dimensional manifold with $n=4$, it follows that the Kretschmann scalar can be decomposed into the traceless Weyl scalar $I$, the quadratic invariant $r_{1}= \widehat{G}_{\alpha}^{\:\:\:{\beta}}\widehat{G}_{\beta}^{\:\:\:{\alpha}}$ (formed from the traceless Einstein curvature tensor), and the square of the Einstein curvature scalar $G$ as follows:

\begin{equation}
\begin{aligned}
K &= I+2 \:{r}_{1} +\frac{1}{6} {G}^2  \quad \text{for } n = 4.\\
\end{aligned}
\label{KIG}
\end{equation}\par

This quadratic invariant decomposition of the Kretschmann scalar can be interpreted (considering free fall geodesic motion) as decomposing the deformation into volumetric and shear modes of deformation. \par
\par
The square of the trace of the Einstein curvature $G^2$ (or equivalently of the trace of the Ricci curvature $R^2$), for a fluid under isotropic pressure, tells us about the curvature due to volume change of spacetime. For example, consider the deformation of an initially perfect sphere changing its radius while remaining spherical. If the pressures are unequal (anisotropic pressure), then, in addition, it contains information about the curvature due to pure shear strain change of spacetime (shear without rotation of the principal axes of stress), in other words, unequal curvature change in different directions. For example, consider the deformation of an initially perfect sphere deforming into an oblate or prolate spheroid.\par

The quadratic traceless invariants, $r_{1}$ and $I$, contain information about the curvature due to simple shear change of spacetime (shear due to planar rotation of the principal axes of stress). There are two possible sources of curvature due to shear change: a local and a nonlocal cause.\par

The invariant $r_{1}$ represents the curvature due to simple shear change arising from local sources, from the self-contraction of the traceless curvature tensor $\widehat{G}_{\alpha}^{\:\:\:{\beta}}\widehat{G}_{\beta}^{\:\:\:{\alpha}}$. This is locally sourced, by virtue of the Einstein field equations, by the traceless stress-energy-momentum tensor $\widehat{T}_{\alpha \beta}$. \par

The invariant $I$ indicates the curvature due to simple shear change due to a non-local source, as represented by the components of the traceless Weyl tensor $C_{\alpha\beta\gamma\delta} C^{\alpha\beta\gamma\delta}$. This is the “free gravitational field” sourced by a distant matter-energy source. A ``Ricci-flat'' space (with $R=0$ and $\widehat{R}_{\alpha \beta}=0$) can still be curved, because vacuum spacetime can be warped. That information is contained in the Weyl tensor. For example, when there is no ordinary matter in spacetime, there can still be gravitational waves, or there can still be gravitational spacetime distortion, which is distantly sourced, for example, from a planet or star.\par
We will consider one cubic invariant formed from the trace-adjusted Ricci curvature tensor (or, equivalently, in four-dimensional spacetime, from the traceless Einstein curvature tensor) as follows:

\begin{equation}
\begin{aligned}
r_{2} &\equiv \widehat{R}_{\alpha}^{\:\:\:{\beta}}\widehat{R}_{\beta}^{\:\:\:{\gamma}} \widehat{R}_{\gamma}^{\:\:\:{\alpha}}\\
 &= \widehat{G}_{\alpha}^{\:\:\:{\beta}}\widehat{G}_{\beta}^{\:\:\:{\gamma}} \widehat{G}_{\gamma}^{\:\:\:{\alpha}}\quad \text{for } n = 4.
\end{aligned}
\label{r2}
\end{equation}\par

Carminatti et al. \cite{Carminati} define this cubic invariant $r_{2}$ with an arbitrary prefactor of $-1/8$. This invariant contains information about the curvature due to simple shear deformation of the purely spatial components, involving rotation of the principal axes of stress in the three-dimensional embedded space.
Comparing the $r_{1}$ and $r_{2}$ invariants, the quadratic invariant $r_{1}$ contains information about the curvature due to simple shear deformation of the purely spatial components, due to the rotation of the principal stress axes on a plane (for example, in-plane strain). The cubic invariant $r_{2}$ contains information about the curvature due to three-dimensional shear deformation of the purely spatial components, which is a more complex state of curvature deformation than the one expressed by $r_{1}$.\par

\subsection{Revisiting Curvature Invariants in the Alcubierre warp-drive}\label{Revisiting}

To the best of our knowledge, Mattingly et al. \cite{Mattingly} appear to be the only authors who have published a study on the curvature invariants of the Alcubierre warp-drive spacetime.

This paper contributed to the understanding of the Alcubierre warp-drive, particularly in the context of curvature invariants. However, it is marred by inconsistencies in the use of units and the presentation of graphical data. \par

In Mattingly et al. \cite{Mattingly}, Figs. 1 and 2, the coordinate dimensions are explicitly denoted as $x\:[m]$ and $t\:[s]$, and velocities are directly stated to be in units of $[m/s]$, despite the authors previously stating their use of natural units in their associated formulas. Contrary to the text's assertion that the warp bubble's velocity behaves as \(v = x/t\) for an Eulerian observer, a simple division of the increments in $x$ by $t$ from their Figs. 1 and 2 yields a warp bubble velocity near $1\;[m/s]$. This is in stark contrast to the figure captions, which specify plotted velocities ranging from 1 to 5 times the speed of light (\(c = 299,792,458\:[m/s]\) ). This discrepancy exceeds eight orders of magnitude. 

In the natural units system where $c=1$, both space and time dimensions should be expressed in consistent units, ensuring that the spacetime interval $ds^2$ is homogeneous in terms of dimensionality. This means that all terms within $ds^2$ should have the same dimensionality, for example, all terms in $ds^2$ should be expressed in units of length (e.g. $[m^2]$), time (e.g. $[s^2]$), or dimensionless. It seems the authors presented plots calculated on the basis of natural units ($c=1$) but labeled them with inconsistent units denoted as $x\:[m]$ and $t\:[s]$. To rectify this, they could have either:
a) adjusted the time label in their figures to represent units of $[m]/c$, equivalent to \(3.3356 \times 10^{-9}\:[s]\) instead of $1\:[s]$, or 
b) recalculated the plots using a time increment of $1\:[s]$, which, given the magnitude of \(c\), would result in a significantly larger $x$ value than depicted. \par

\begin{figure}[ht]
    \centering
    \includegraphics[width=0.9\textwidth]{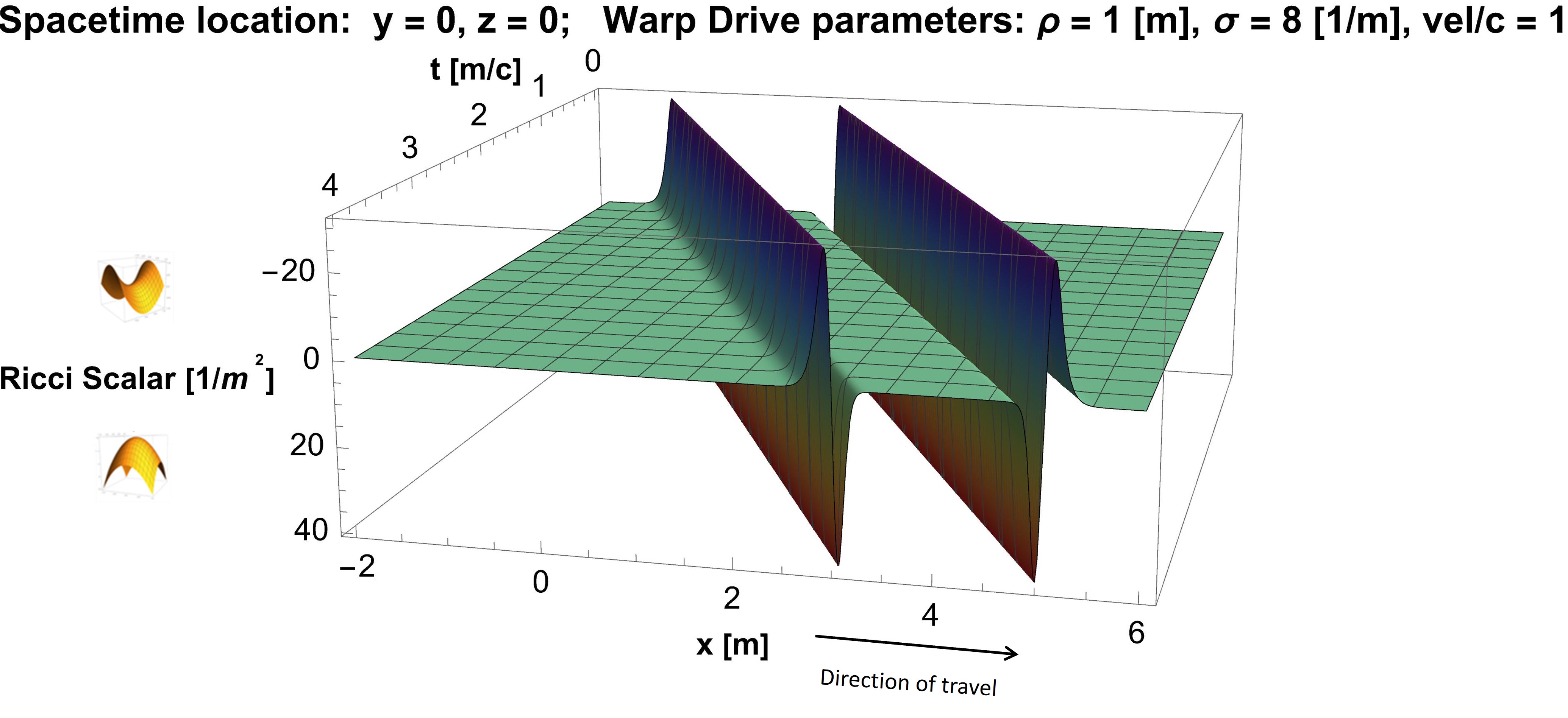}
    \caption{3-D plot of the Ricci scalar invariant $R\:[1/m^2]$ vs. $x\:[m]$ and $t\:[m/c]$, for the warp-drive with the parameters Alcubierre used in his paper \cite{Alcubierre_1994}. The spaceship is traveling at a constant velocity (the speed of light $c$) in the positive $x$ direction, as seen by Eulerian distant external observers.}
    \label{Fig0_Ricci}
\end{figure}

In Fig.~\ref{Fig0_Ricci}, we present a 3-D plot of the Ricci scalar invariant $R$ for the Alcubierre warp-drive against $x\:[m]$ and $t\:[m/c]$, using the parameters specified by Alcubierre \cite{Alcubierre_1994}. This plot depicts a spaceship moving at a constant velocity, equivalent to the speed of light $c$, in the positive $x$ direction, as observed by distant Eulerian observers. This plot serves as an example of correct unit representation: the time label reflects units of $[m/c]$, equivalent to $3.3356 \times 10^{-9} \: [s]$. A simple division of the increments in $x$ by $t$ in Fig.~\ref{Fig0_Ricci} yields the correct warp bubble velocity. 
 Also, contrary to the approach of Mattingly et al. \cite{Mattingly}, our plot does not truncate the magnitude of $R$ arbitrarily at an extremely small amplitude. Instead, our Fig.~\ref{Fig0_Ricci} accurately represents the true magnitude of the spacetime warping associated with the Alcubierre warp-drive. This underscores the importance of accurate representation in such plots.

The absence of plots illustrating the planar or three-dimensional distribution of the invariants is an oversight. \par

Additionally, the truncation of Ricci's curvature scalar $R$ in the $x$ (direction of travel) vs. $t$ (time) plots in  Fig.1 of Mattingly et al. \cite{Mattingly}, by 8 to 16 orders of magnitude is noteworthy. Specifically, their Fig.1b (for $v_s=1\:c$) is truncated at amplitudes of $\approx 10^{-7} \: [1/m^2]$ and their Fig.1d (for $v_s=4\:c$) is truncated at amplitudes of  $\approx 10^{-15} \: [1/m^2]$. This contrasts with the true amplitude range shown in our Fig.~\ref{Fig0_Ricci} of $\approx -20\:[1/m^2]\; \text{to} +40\:[1/m^2]$ for $v_s=1\:c$. Their truncation appears arbitrary and lacks justification.\par

To understand the importance of accurately representing curvature magnitude, it is beneficial to compare it with the spacetime curvature of a black hole. For the exterior vacuum of a Schwarzschild black hole, the Ricci scalar $R$ is zero, but the Kretschmann scalar (Eq.(\ref{Kretschmann},\ref{KIG})) is not. The Kretschmann scalar for such a black hole is given by $K =12\: r_s^2/r^6 $, where $r_s$ represents the Schwarzschild radius, also known as the event horizon radius. When evaluating the square root of the Kretschmann scalar at the event horizon radius $r=r_s$, the resulting curvature is $\sqrt{K}=3.464101/r_s^2$. This curvature measure is inversely proportional to the square of the radial distance of the event horizon. Hence, the smaller the Schwarzschild radius, the higher the spacetime curvature as measured by the square root of the Kretschmann scalar at that location.\par 

In our analysis (Fig.~\ref{Fig0_Ricci}), the peak amplitude of the Ricci scalar for the Alcubierre warp-drive traveling at the speed of light ranges from $\approx -20 \: [1/m^2]$ to $\approx +40 \: [1/m^2]$. This is 4 to 8 times greater than the curvature measured with the square root of the Kretschmann scalar $\sqrt{K} = 4.863 \: [1/m^2]$ at the event horizon of a Schwarzschild black hole with the mass equivalent of the planet Saturn, which has an event horizon radius of $r_s=0.844\; [m]$.\par

For context, the $10^{-7} \: [1/m^2]$ value plotted in Fig.1b of Mattingly et al. \cite{Mattingly} corresponds to the spacetime curvature $\sqrt{K}$ at a radial distance of $291\:[m]$, which is $345$ times the Schwarzschild radius $r_s=0.844\; [m]$ for the same black hole. The smaller value of $10^{-9} \: [1/m^2]$ discussed in the text by Mattingly et al. \cite{Mattingly} represents the curvature $\sqrt{K}$ at a radial distance of $1351\:[m]$, $1601$ times the Schwarzschild radius. The much smaller value of $10^{-15} \: [1/m^2]$ plotted in Fig.1d of Mattingly et al. \cite{Mattingly} represents the curvature $\sqrt{K}$ at a radial distance of $135132\:[m]$, $160109$ times the Schwarzschild radius.\par

The theoretical black hole, used here for comparison with the spacetime curvature of the Alcubierre warp-drive, has a mass equivalent to that of the planet Saturn. This makes it notably smaller than a stellar mass, distinguishing it from the much more massive black holes that have been observed astronomically. Such smaller black holes, with masses less than stellar values, may have originated in the high-density conditions of the early universe, and are referred to as primordial black holes.  Over time, these primordial black holes would either grow by accreting surrounding matter or diminish due to the effects of Hawking radiation.\par

Since Mattingly et al. \cite{Mattingly} write, ``This can be seen as $\sigma \rightarrow  8\: [1/m]$, as the Ricci scalar goes to $10^{ -9}$, $r_{1}$ goes to $10^{ -11}$, and $w_{2}$ goes to the order of $10^{ -28}$,'' it seems that the authors might not have been aware of a default behavior in the Wolfram \textit{Mathematica\textsuperscript{\textregistered}} software they used to plot their calculations. By default, \textit{Mathematica} auto-adjusts, and therefore truncates the range of plots to emphasize regions where the function exhibits the most varied behavior. With the default setting of \texttt{PlotRange $\rightarrow$ Automatic}, \textit{Mathematica} often chooses not to show the highest or lowest values. The truncation of the Ricci scalar's amplitude by 8 to 16 orders of magnitude in the plots of Mattingly et al. \cite{Mattingly} is indicative of this behavior. The displayed Ricci scalar in their figures exhibits clear signs of this truncation, characterized by a flat white surface parallel to the base plane of the 3-D plot.  This is consistent with \textit{Mathematica}'s default representation for regions where the actual amplitude of the plot has been clipped by automatically adjusting the plotted range. As a result, the diminutive amplitudes Mattingly et al. present in their plots and discuss in their text are many orders of magnitude below the true amplitude of the Ricci scalar $R$. When discussing the Alcubierre warp-drive concept, it is imperative to accurately represent the magnitude of spacetime curvature changes. Inaccurately representing by many orders of magnitude, as done by the authors, the substantial curvature change that the Alcubierre warp-drive concept entails could lead to misconceptions about its feasibility and implications. The values they mention appear to be the outcomes of \textit{Mathematica}'s automatic range adjustments. To ensure an accurate representation of the function’s amplitude, it is essential to set the desired range in \textit{Mathematica}, setting \texttt{PlotRange $\rightarrow$ All}.
 This approach ensures that the visualization is both complete and free from potential misinterpretations.

\subsubsection{Clarifying the Classification of the Alcubierre Spacetime}\label{Clarifying}

It is essential to note that the assertion (made without proof) by Mattingly et al. \cite{Mattingly}  that: ``For Class $B_{1}$ spacetimes, which include all hyperbolic spacetimes, such as the general warp-drive line element, only four CM invariants, $R, r_{1}, r_{2},$ and $w_{2}$, are necessary to form a complete set'' is invalid, because the Alcubierre warp-drive line element is not a Class $B_{1}$ Warped Product Spacetime. 

Their assertion \cite{Mattingly} that the cubic invariant $r_{2}$ vanishes for the Alcubierre metric is also incorrect. These discrepancies hinder the paper's \cite{Mattingly} clarity and ability to effectively communicate its findings.

In reference to the assertions made in \cite{Mattingly}, it is pertinent to clarify that the notions of a globally hyperbolic spacetime and a Class $B$ spacetime represent distinct characteristics within general relativity. Being globally hyperbolic is not a sufficient requirement for a spacetime to be of Class $B$.\par

\textit{Definition (Globally Hyperbolic):}  A spacetime is termed \textit{globally hyperbolic}  \cite{hawking_ellis_1973} if it contains a \textit{Cauchy surface}, which is a hypersurface intersected precisely once by every inextendible, non-spacelike curve.  Such a curve delineates the complete worldline of a particle (or a light ray) from its origin to its termination within the spacetime, uninterrupted by boundaries or singularities and without representing superluminal motion.  The entire past and future of the whole spacetime can be either predicted or retrodicted just from knowledge of the Cauchy surface. Global hyperbolicity is a condition on the causal structure of a Lorentzian manifold, and it implies that the spacetime excludes the presence of certain types of singularities or ``edges'' at infinity within the spacetime. This condition ensures determinism in general relativity, as initial conditions on this surface uniquely determine the spacetime's evolution. \par

\textit{Definition (Class $B$ Warped Product Spacetimes):} \textit{Class B warped product spacetimes} \cite{Carot_1993,Carot2005,Santosuosso} are spacetimes expressible as the product of two distinct 2-dimensional spaces: one with a Lorentzian metric and the other with a Riemannian metric. The coupling between these spaces should be separable. The Class $B$ warped product spacetime classification is based on the metric's mathematical structure rather than its causal properties, thus this classification is separate and distinct from the condition of the spacetime being globally hyperbolic. Not all globally hyperbolic spacetimes fall under the category of Class $B$ warped product spacetimes.\par

The canonical form of a class $B_{1}$ warped product spacetime is:

\begin{equation}
\begin{aligned}
ds^2 &= -2\: f(u,v)\: du\: dv + r(u,v)^2\:g(\theta, \phi)^2\: (d\theta^2 + d\phi^2).\\
\end{aligned}
\label{B1Def}
\end{equation}\par

and the standard form for class $B_{2}$ warped product spacetime is:

\begin{equation}
\begin{aligned}
ds^2 &= f(u,v)^2\: (du^2+dv^2) - 2\:r(u,v)^2\:g(\theta, \phi)\: d\theta \: d\phi.\\
\end{aligned}
\label{B2Def}
\end{equation}\par

\subsubsection{Analysis of the Alcubierre Metric} \label{sec:analysis_alcubierre}

The prototypical form of the Alcubierre metric \cite{Alcubierre_1994,AlcubierreLobo} (defined for velocity $v(t)$ along the $x$ direction), is:

\begin{equation}
\begin{aligned}
ds^2 &= -dt^2+(dx-v(t)\:f(x-x_{0}(t),y,z)\:dt)^2+dy^2+dz^2.\\
\end{aligned}
\label{Alcubierre}
\end{equation}\par

Alcubierre's line element is based on the 3+1 (or ADM) formalism, which decomposes spacetime into spacelike hypersurfaces labeled by a time parameter. Each hypersurface is positive definite by construction, ensuring real and positive spatial distances on each individual \textit{local} hypersurface. However, this doesn't preclude the \textit{global} Alcubierre spacetime from having closed timelike curves. The 3+1 formalism ensures a specific causal structure \textit{locally} but doesn't guarantee \textit{global} hyperbolicity.  A classic example of a spacetime that is locally causal but has a global noncausal structure is the Gödel metric \cite{OzsváthSchücking}. Ralph et al. \cite{SpinAlcubierre} present a global metric representing an Alcubierre warp-drive on a rotating platform. It describes closed timelike curves embedded in Minkowski spacetime. 

The eigenvalues of the Alcubierre metric are:

\begin{equation}
\label{eigen}
\begin{aligned}
\begin{bmatrix}
    \lambda_1 \\
    \lambda_2 \\
    \lambda_3 \\
    \lambda_4 \\
\end{bmatrix}
=
\begin{bmatrix}
    \frac{1}{2} (v(t)^2 f(x-x_{0}(t),y,z)^2-\sqrt{4+v(t)^4 f(x-x_{0}(t),y,z)^4})\\
    \frac{1}{2}(v(t)^2 f(x-x_{0}(t),y,z)^2+\sqrt{4+v(t)^4 f(x-x_{0}(t),y,z)^4})\\
    1 \\
    1 \\
\end{bmatrix}.
\end{aligned}
\end{equation}

Given these eigenvalues, the Alcubierre metric exhibits a Lorentzian signature metric with signature $(-,+,+,+)$, characterized by one negative eigenvalue ($\lambda_1$) corresponding to a timelike direction, and three positive eigenvalues ($\lambda_2,\lambda_3,\lambda_4$) corresponding to spacelike directions, for any arbitrary value of the ship's velocity $v(t)$ times the form function $ f(x-x_{0}(t),y,z)$.

When juxtaposed with the typical form of a class $B_{1}$ Eq.(\ref{B1Def}) or a class $B_{2}$ Eq.(\ref{B2Def}) warped spacetime, it becomes evident that the Alcubierre metric Eq.(\ref{Alcubierre}) does not conform to these Class $B$ spacetime structures. The problem is the coupling term due to the form function $f(x-x_{0}(t),y,z)$ that intertwines all the spacetime coordinates $\{t,x,y,z\}$, including the time coordinate, since the center of the ship's position $x_{0}(t)$ is a function of time $t$. The form function $f$ in the Alcubierre metric, which depends on $x,x_{0}(t),y,z$ introduces dependencies that prevent a straightforward decomposition of the Alcubierre metric into two independent 2-dimensional spaces, one Lorentzian and the other Riemannian.\par

\begin{figure}[h!]
    \centering
    \begin{subfigure}[b]{0.3\textwidth}
        \includegraphics[width=\textwidth]{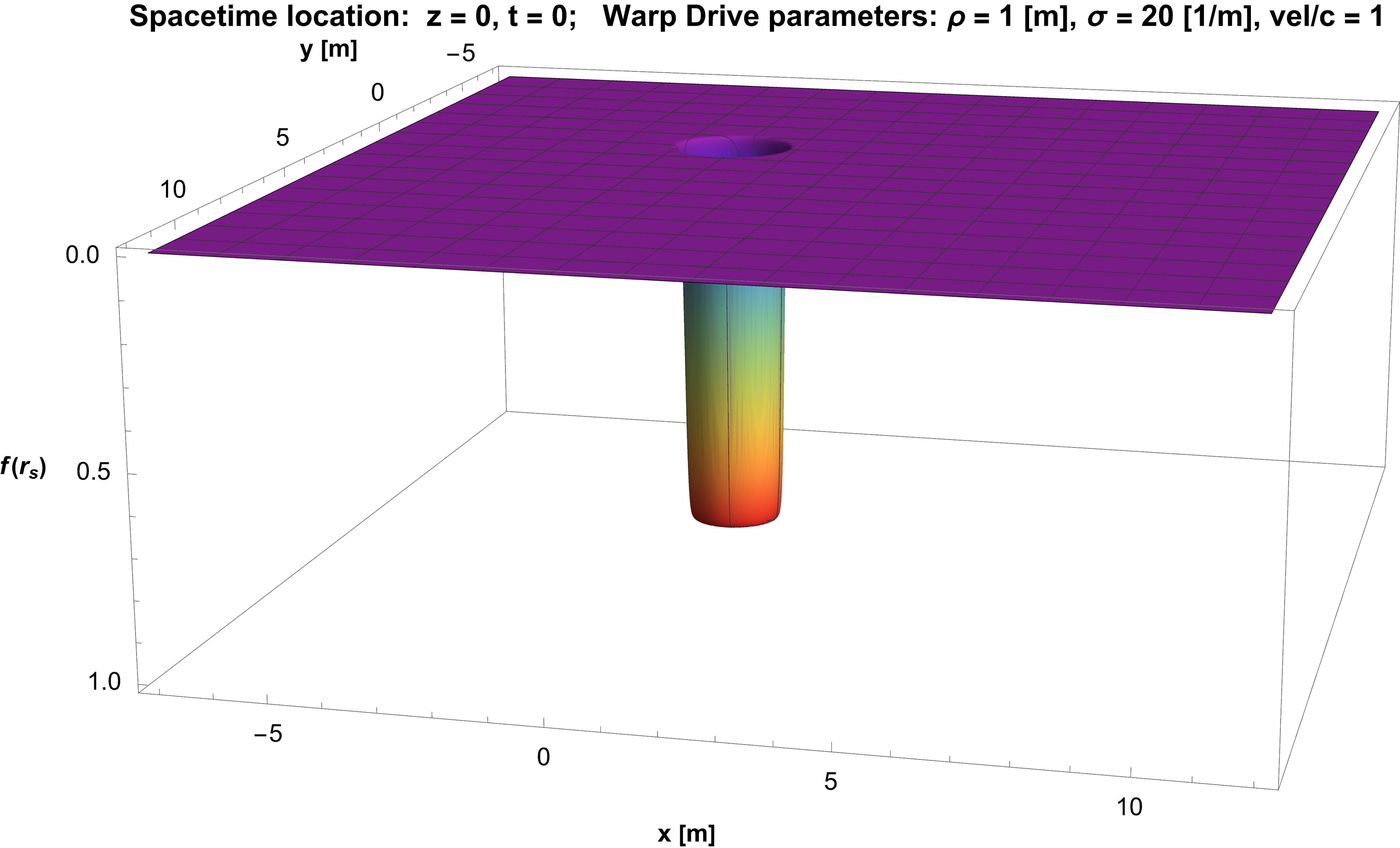}
        \caption{$\sigma=20\:[1/m]$}
        \label{fig:a}
    \end{subfigure}
    \hfill
    \begin{subfigure}[b]{0.3\textwidth}
        \includegraphics[width=\textwidth]{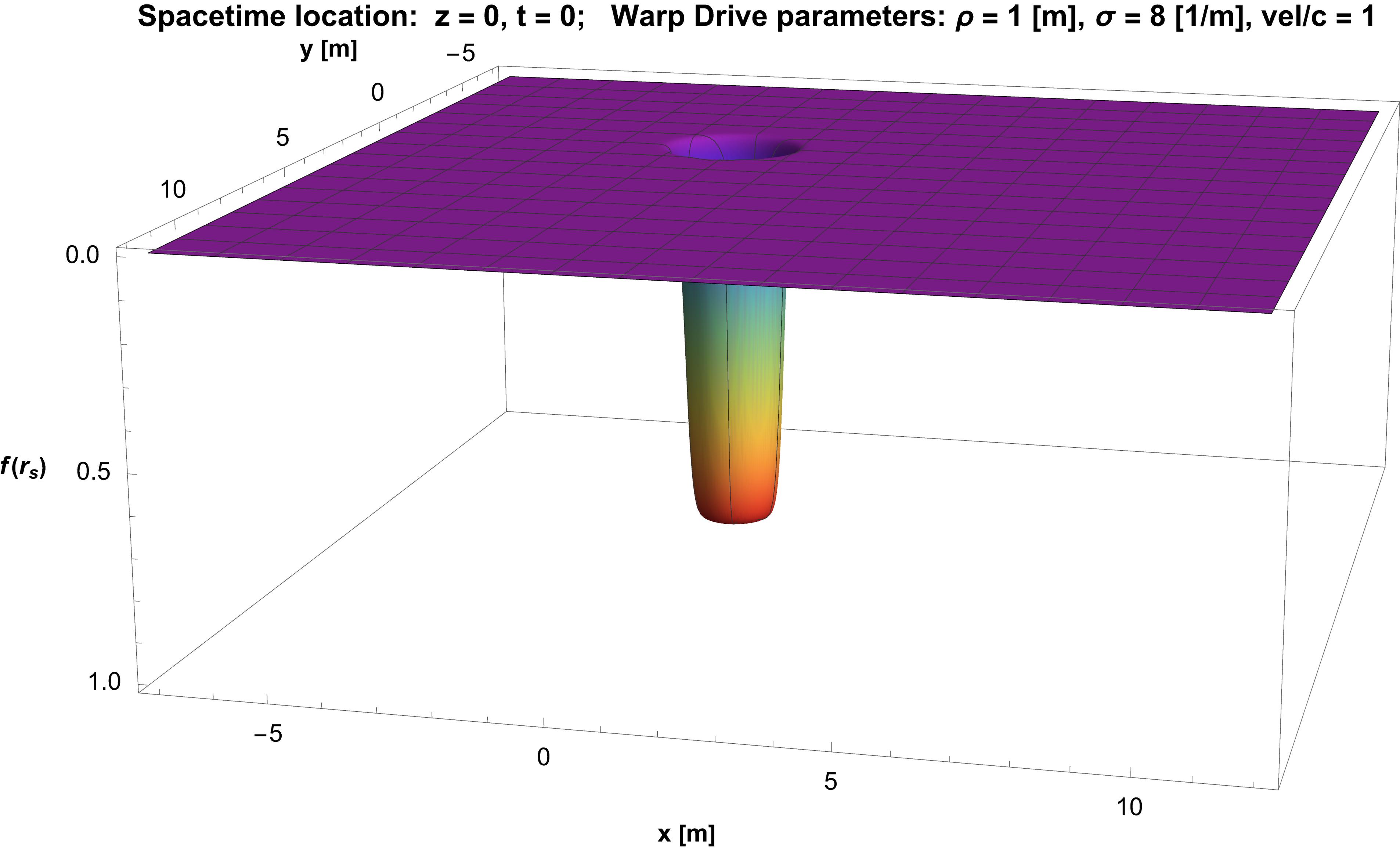}
        \caption{$\sigma=8\:[1/m]$}
        \label{fig:b}
    \end{subfigure}
    \hfill
    \begin{subfigure}[b]{0.3\textwidth}
        \includegraphics[width=\textwidth]{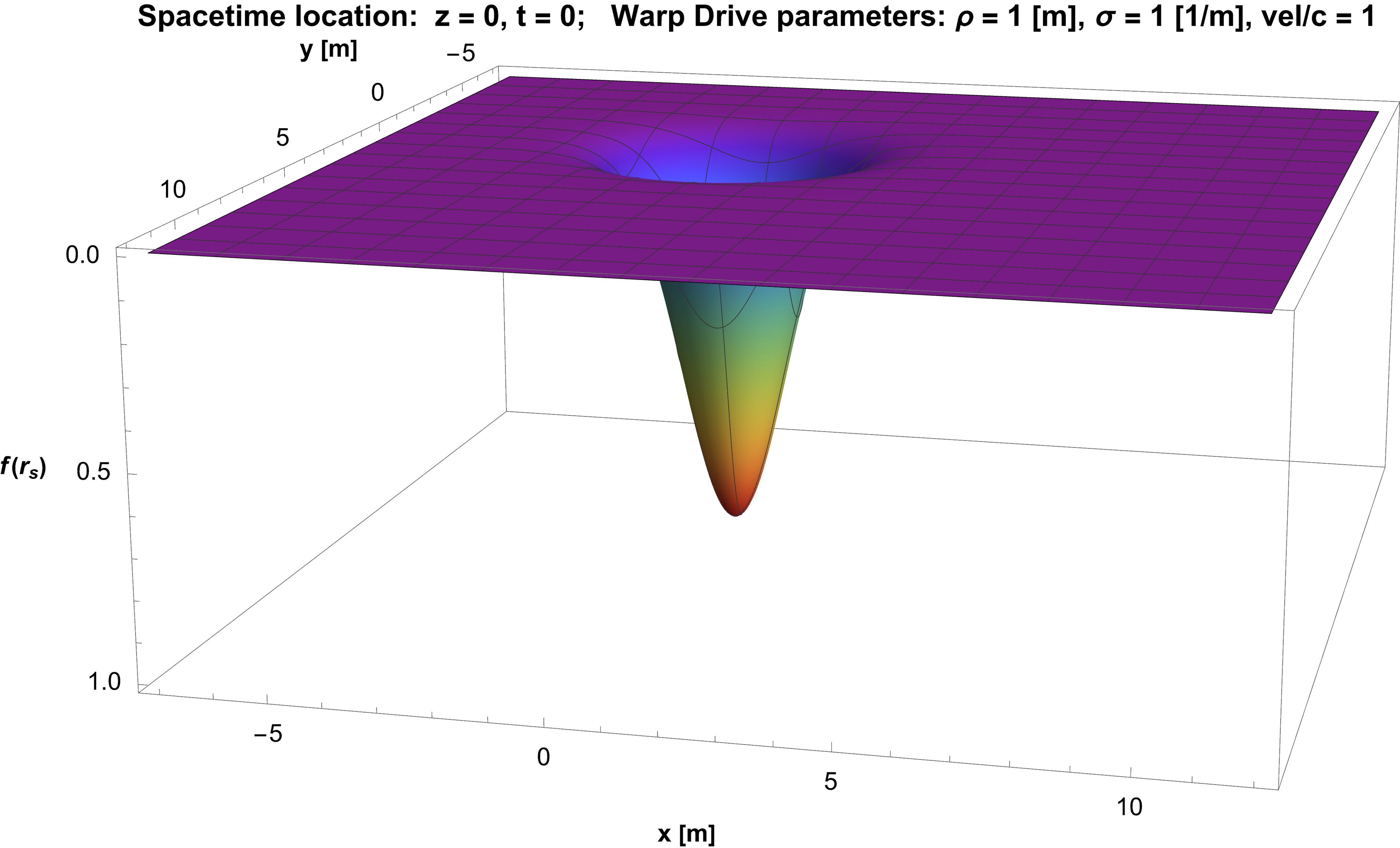}
        \caption{$\sigma$=1\:[1/m]}
        \label{fig:c}
    \end{subfigure}
    \caption{``Top hat'' shape of Alcubierre's form function Eq.(\ref{AlcubierreForm}) plotted as a function of the plane coordinates $x$ and $y$, (at $t = 0$ and $z = 0$), for different values of the warp bubble's `inverse thickness' parameter $\sigma$, at constant warp bubble `radius' $\rho=1\: [m]$. The spaceship is traveling at a constant velocity (the speed of light $c$) in the positive $x$ direction, as seen by Eulerian distant external observers.}
    \label{TopHat}
\end{figure}

In particular, the specific form function, $f(x-x_{0}(t),y,z)$, chosen by Mattingly et al. \cite{Mattingly} was the same ``top hat'' (Fig.~\ref{TopHat}) symmetric form chosen by Alcubierre:

\begin{equation}
\begin{aligned}
f(x-x_{0}(t),y,z) &=\frac{\tanh{[\sigma (r(t,x,y,z)+\rho)}]-\tanh{[\sigma (r(t,x,y,z)-\rho)}]}{2\tanh{[\sigma \rho]}}\\
r(t,x,y,z) &=\sqrt{(x-x_{0}(t))^2+y^2+z^2}.
\end{aligned}
\label{AlcubierreForm}
\end{equation}\par

Evidently, all coordinates are intertwined in the transcendental function $\tanh{[\sigma (\sqrt{(x-x_{0}(t))^2+y^2+z^2}+\rho)}]-\tanh{[\sigma (\sqrt{(x-x_{0}(t))^2+y^2+z^2}-\rho)}]$ which is not separable into Lorentzian and  Riemannian distinct 2-dimensional spaces.\par

In conclusion, the form of the Alcubierre metric chosen by Mattingly et al. \cite{Mattingly}, though showcasing warping of spacetime, does not meet the specific criteria \cite{Carot_1993,Carot2005,Santosuosso} set for ``class $B$'' warped spacetimes.

% REQUESTED MINOR REVISION TEXT (OCT 16 2023) BEGINS HERE

Carneiro et.al. \cite{Carneiro2022} stated (without proof) that the Petrov classification for the Alcubierre spacetime is Petrov type I. This is the most general Petrov type: it does not possess any algebraic symmetry and has four real, simple, and unequal principal null directions. We confirm that the Alcubierre spacetime is indeed Petrov type I based on the following facts:

\begin{enumerate}
    \item The five complex Newman-Penrose scalars for the Alcubierre metric (calculated based on the null vectors of \cite{Mattingly}) are all unequal and non-zero ($\Psi_{0}, \Psi_{1}, \Psi_{3}, \Psi_{4}$ are all complex, and $\Psi_{2}$ is purely real). This indicates Petrov type I \cite{stephani2003}. The complex nature of the Newman-Penrose scalars $\Psi_{0}, \Psi_{1}, \Psi_{3}, \Psi_{4}$ is due to the participation of both the electric (tidal deformation) and magnetic (twisting of null geodesic congruences) parts of the Weyl tensor \cite{Hofmann2013}. 
    
    \item The 3x3 complex, traceless matrix of $Q^{\alpha}\,_{\beta} = E^{\alpha}\,_{\beta} + i \, H^{\alpha}\,_{\beta} $, which encodes the principal null directions \cite{stephani2003}, has three distinct, unequal eigenvalues $\lambda_{1} \neq \lambda_{2} \neq \lambda_{3}$. $E^{\alpha}\,_{\beta}$ is the electric part, and $H^{\alpha}\,_{\beta}$ the magnetic part of the Weyl tensor. The Segre characteristic is $[111]$ since each of the three eigenvalues has a multiplicity of 1.  Therefore the spacetime must be Petrov type I \cite{stephani2003}. 
    
    \item It is known \cite{stephani2003, Bini2023} that Petrov type I is the only spacetime that satisfies the inequality $\widetilde{I}^{3} \neq  27 \: \widetilde{J}^{2}$. In this inequality, $\widetilde{I}\coloneqq \frac{1}{32} Q^{\alpha}\,_{\beta} \; Q^{\beta}\,_{\alpha}$ is the quadratic complex curvature scalar invariant and $\widetilde{J}\coloneqq \frac{1}{384} Q^{\alpha}\,_{\beta} \; Q^{\beta}\,_{\delta} \; Q^{\delta}\,_{\alpha}$ is the cubic complex curvature scalar invariant.  Calculation shows that that for the Alcubierre spacetime, $\widetilde{I}^{3} \neq  27 \: \widetilde{J}^{2}$, therefore the spacetime must be Petrov type I \cite{stephani2003, Bini2023}.  
\end{enumerate}

Class B warped spacetimes can only be Petrov type D or O \cite{Carot_1993}. Therefore, since the Alcubierre spacetime is Petrov type I, it cannot be a Class B warped spacetime.

In closing our discussion on symmetries and spacetime classifications, it is pertinent to note that the Alcubierre warp drive can be analogously visualized \cite{Baak2023} within a spacetime framework as a soliton moving inside a fluid. The spherical symmetry pertains to the geometrical shape of the warp bubble, not to the enveloping fluid flow. The analog gravity \cite{Baak2023} streaming flow around the spherical warp bubble exhibits the most general type: Petrov type I. 

% REQUESTED MINOR REVISION TEXT (OCT 16 2023) ENDS HERE

\section{Visualizing Invariants in the Alcubierre warp-drive}

\subsection{Representation and Interpretation of the Alcubierre Warp Bubble}
% Content for the representation and interpretation of the Alcubierre Warp Bubble.

\begin{figure}[ht]
    \centering
    \includegraphics[width=0.9\textwidth]{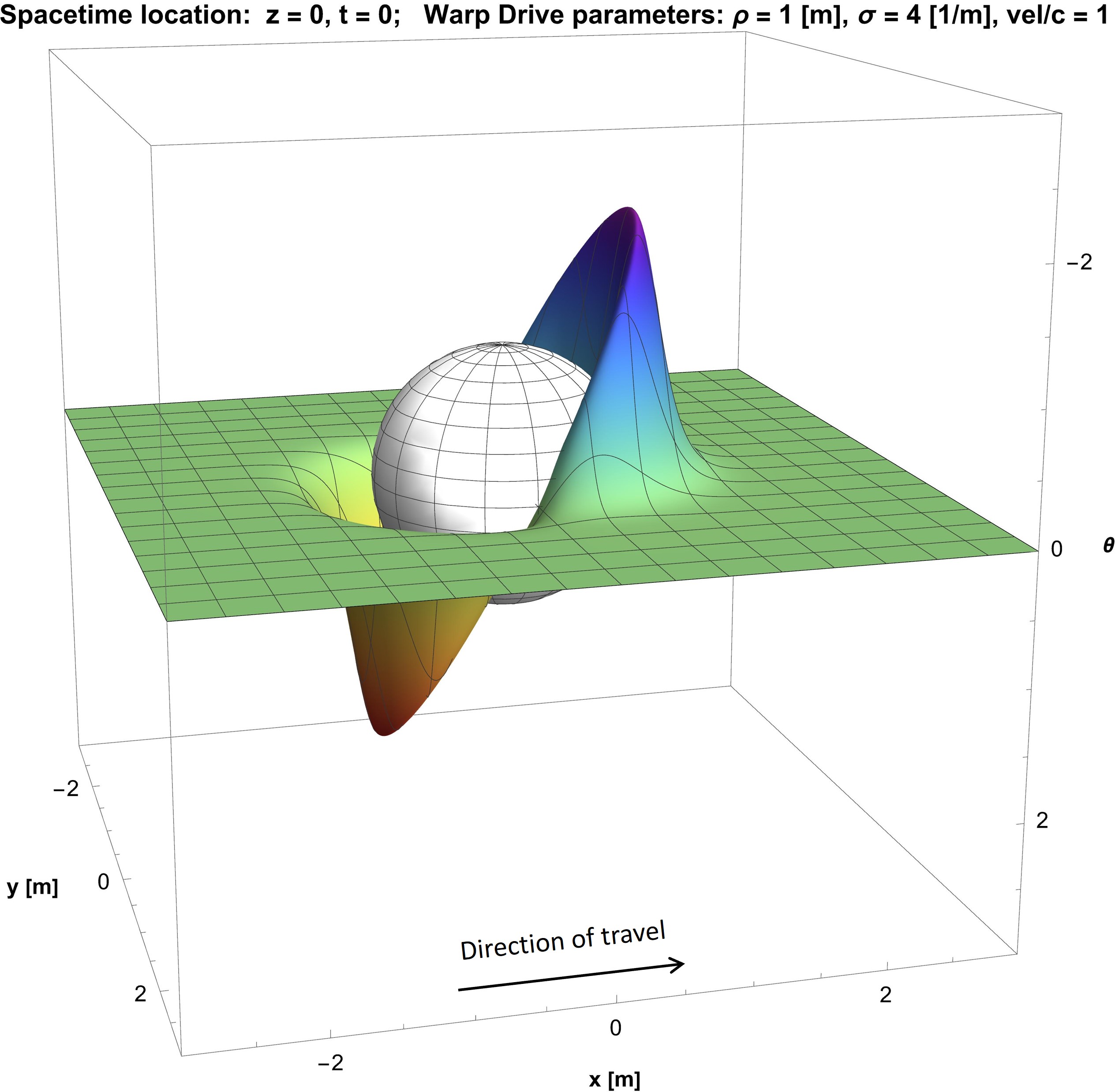}
    \caption{Expansion and contraction of volume elements ($\theta$) of the Alcubierre metric
associated with external Eulerian observers and the spherical fuselage that matches the Alcubierre warp-bubble contour.}
    \label{Fig1_Theta}
\end{figure}

The seminal paper by Alcubierre \cite{Alcubierre_1994} considers the warp-drive as a test particle, located at a point ($x_{0}(t)$) at the center of the warp bubble. This test particle assumption idealizes the spaceship by assuming its mass to be so insignificant that the spaceship's trajectory is a geodesic. Alcubierre's proposal was purely theoretical and was based on the mathematics of general relativity. He did not provide a mechanism for creating such a warp bubble, but he noted that if it were possible, it would require ``exotic matter'' with ``negative energy density.'' The feasibility of creating and controlling such exotic matter, as well as the potential effects and consequences of using such a drive, are still topics of ongoing research and debate in the scientific community. However, the test particle assumption does not imply that the spaceship should be depicted inside the warp bubble as if it has no influence on determining the warp bubble. Consider the following analogies: in Newtonian mechanics, the Earth's gravitational field can be thought of as emanating from a point particle situated at the center of the Earth, with all of the Earth's mass concentrated at that point. In Einstein's general relativity, the Schwarzschild solution provides the most general spherically symmetric vacuum solution to the Einstein field equations. This solution describes a Schwarzschild black hole, a static black hole with no electric charge or angular momentum. Such a black hole is uniquely defined by its mass, and features an event horizon at the Schwarzschild radius. However, it would be a misrepresentation to depict the Earth's gravitational field as if the Earth were merely a point or had a radius of just $9\:[mm]$ (equivalent to the Earth's Schwarzschild radius). In reality, the Earth's gravitational pull is most pronounced at its surface, approximately $6400\:[km]$ from its center.
Similarly, the warp bubble is produced by the medium situated within the ship's fuselage. The peak of the warp bubble's field should align with the fuselage of the warp-drive, analogous to how the Earth's maximum gravitational attraction occurs at its surface.
The assumption of a test particle involves neglecting factors such as the rotation of the warp-drive and any other elements that could cause the ship's trajectory to deviate from a geodesic, due to the ship’s finite size. This can be addressed by depicting a straight geodesic, rather than separating the spaceship from the warp bubble's size in the illustration.
In his paper \cite{Alcubierre_1994}, Alcubierre did not illustrate any invariant. Instead, he illustrated the expansion $\theta$ of volume elements associated with external Eulerian observers, showing volume elements expanding behind the spaceship, and antisymmetrically contracting in front of it. It is important to note that this is a measure of extrinsic curvature, not intrinsic curvature. Unlike the Ricci or the Einstein curvature scalars, which are invariants and a measure of intrinsic curvature, the expansion of volume elements measured by the extrinsic curvature is not an invariant. It depends on the choice of observers and their trajectories, reflecting the extrinsic nature of the geometry. Since the publication of Alcubierre's paper \cite{Alcubierre_1994}, most subsequent authors have reproduced his plot. This approach contrasts with using an invariant measure of intrinsic curvature in this report, which remains constant regardless of the observer's perspective or coordinate system, and has a symmetric distribution fore and aft of the warp-drive.
White \cite{WhiteGRG,White101,White102} depicted the expansion and contraction of volume elements, as described in Alcubierre's paper \cite{Alcubierre_1994}, but also included a spaceship within the depicted bubble. His representation, particularly in the direction of travel, had the warp bubble extending significantly beyond the spaceship in both the fore and aft directions, not aligning with the spaceship's fuselage.  In reality, the Alcubierre bubble must be generated by the spaceship itself utilizing matter-energy. Therefore, the shape of the bubble should match the spaceship's fuselage. The first authors that appear to have pointed this out are Bobrick et al. \cite{Bobrick_2021} and Sarfatti \cite{Sarfatti}. A spacecraft correctly matching Alcubierre’s warp bubble is shown in Fig.~\ref{Fig1_Theta} of the expansion and contraction of volume elements $\theta$ associated with external Eulerian observers, vs. the $x,y$ coordinates, with the spacecraft traveling at a constant velocity (the speed of light $c$) in the positive $x$ direction.

\subsection{\texorpdfstring{Einstein’s Curvature Scalar Invariant, \( {G} \)}{Einstein's Curvature Scalar G}}
% Discussion and plots related to Einstein’s curvature scalar.

All plots of curvature invariants were calculated and plotted using Wolfram \textit{Mathematica\textsuperscript{\textregistered}} \cite{Mathematica} based on the Alcubierre line element Eq.(\ref{Alcubierre},\ref{AlcubierreForm}). To ensure the accuracy of our calculations, the results of all expressions used to calculate the curvature invariants were cross-verified against a range of known exact solutions. In Appendix~\ref{secA1}, we provide the closed-form analytical expressions for the curvature invariants $G,r_1, I$, and $r_2$. For the purpose of direct comparison, all plots are expressed in terms of a form with consistent curvature units, such as $G,\sign{[r_{1}]}\sqrt{\abs{r_{1}}}, \sign{[I]}\sqrt{\abs{I}}$, and $\sign{[r_{2}]}\sqrt[3]{\abs{r_{2}}}$. It is noteworthy that these invariants, when rendered in this consistent curvature unit form, all manifest a dependence on the square of the velocity, analogous to how energy is proportional to velocity squared.

\par

\begin{figure}[ht]
    \centering
    \includegraphics[width=1.\textwidth]{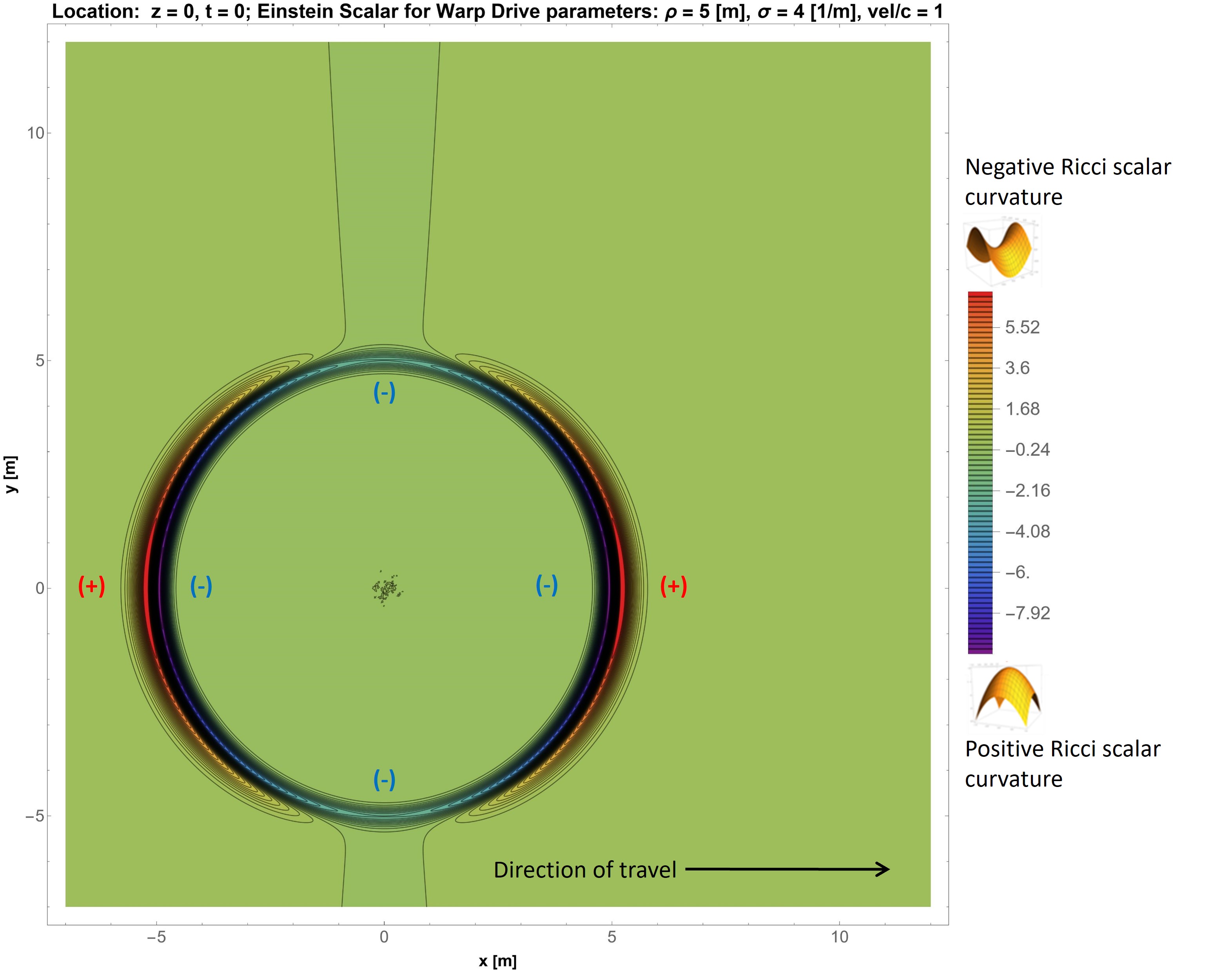}
    \caption{Contour plot in the $x,y$ plane of the intensity of Einstein’s curvature scalar $G= G_{\alpha}\,^{\alpha}$, of the Alcubierre warp bubble. The spaceship is traveling at a constant velocity (the speed of light $c$) in the positive $x$ direction, as seen by Eulerian distant external observers.}
    \label{Fig2_G_Planform}
\end{figure}

We begin with Fig.~\ref{Fig2_G_Planform}, a contour plot (planform view) in the $x,y$ plane (at $t=0$ and $z=0$) of the intensity of Einstein’s curvature scalar $G\equiv tr(Eins)=G_{\alpha}\,^{\alpha}$, using the metric Eq.(\ref{Alcubierre}) of the Alcubierre warp bubble. The Alcubierre parameters are set at (warp bubble ``radius'') $\rho=5\: [m]$, and (warp bubble ``inverse thickness'') $\sigma=4\: [1/m]$. The spaceship is traveling at a constant velocity (the speed of light $c$) in the $x$ direction, as seen by Eulerian distant external observers. 

Fig.~\ref{Fig2_G_Planform} shows that the warp bubble (and hence the fuselage of the spaceship that is responsible for generating Alcubierre’s spherical warp bubble) requires two spherical layers (inner and outer) to be consistent with the distribution of $G$.\par
\textbf{Inner Circumferential Layer}: Surrounding and adjacent to the interior Minkowski flat spacetime, this layer has a negative value of Einstein's curvature scalar ($G<0$). This can be realized for example, with the inner layer having positive energy density ($\rho\:c^2>0$) and zero fluid pressure ($p=0$). This inner layer is also associated with a positive ($R>0$) Ricci curvature scalar (as might occur when an initially Minkowski flat spacetime deforms into a positive curvature hypersurface, as sourced by positive gravitational matter).\par
\textbf{Outer Circumferential Layer}: Adjacent to the exterior spacetime of the warp bubble, this layer has a positive value of Einstein's curvature scalar ($G>0$). This external layer can be realized for example, with negative energy density ($\rho\:c^2<0$) and zero fluid pressure $p=0$. This is the conventional view expressed in many papers, but this is unrealistic as there is no experimental observation of negative energy density, so far, in our universe. A more practical realization is a fluid with sufficiently greater fluid hydrostatic pressure than energy density ($3\:p>\rho\:c^2$). This outer layer is also associated with a negative ($R<0$) Ricci curvature scalar, as might occur when an initially Minkowski flat spacetime deforms into a negative curvature, like a saddle, sourced by positive hydrostatic pressure greater in magnitude than energy density.\par

\begin{figure}[ht]
    \centering
    \includegraphics[width=1.\textwidth]{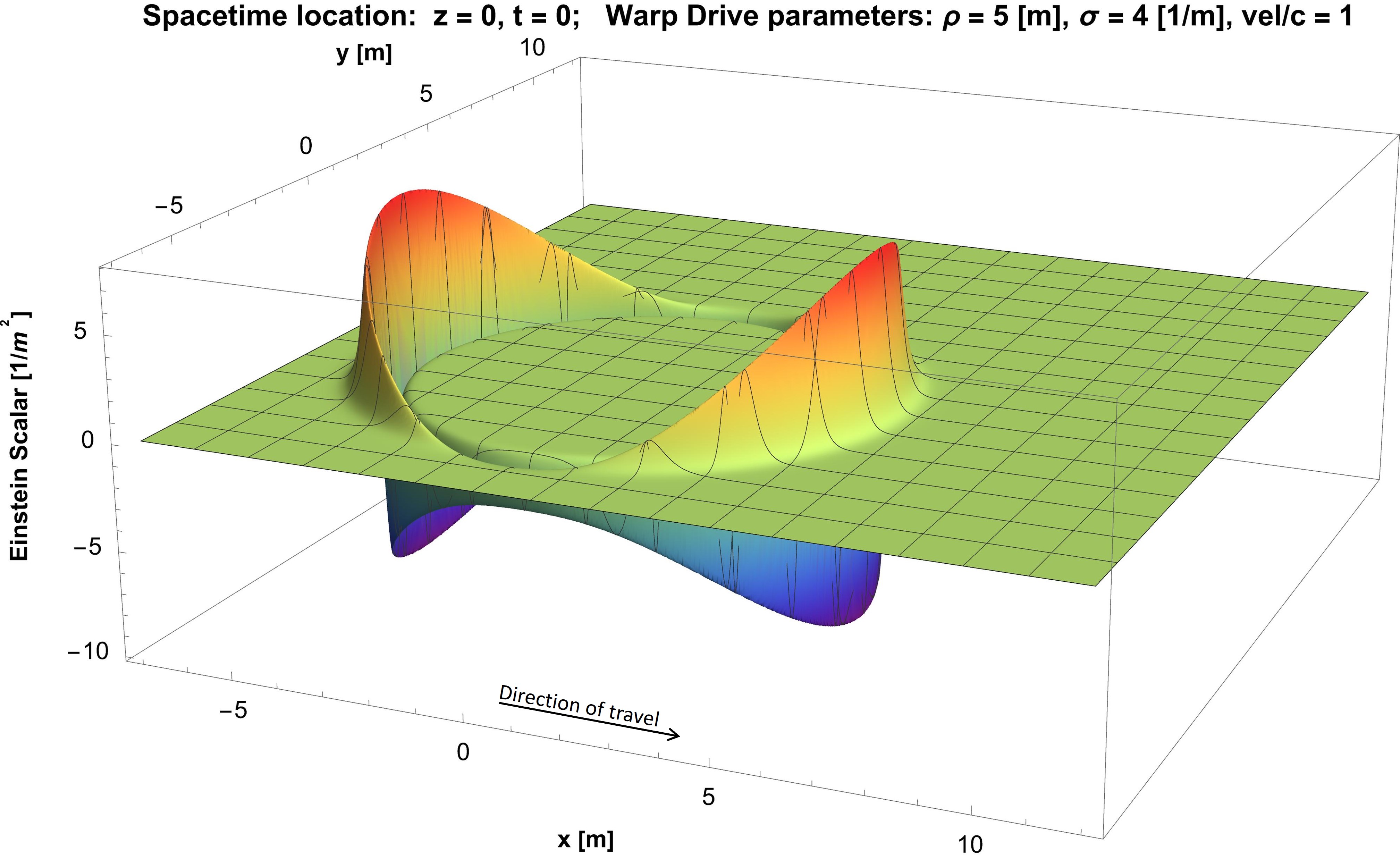}
    \caption{3-D plot of the intensity of Einstein’s curvature scalar $G= G_{\alpha}\,^{\alpha}$, of the Alcubierre warp bubble vs. the $x,y$ coordinates. The spaceship is traveling at a constant velocity (the speed of light $c$) in the positive $x$ direction, as seen by Eulerian distant external observers. A view from above, displaying positive values of Einstein's curvature scalar ($G>0$). The inner Minkowski flat portion of the spaceship, where $G=0$, is clearly shown.}
    \label{Fig3_G_3D_plus}
\end{figure}

\begin{figure}[ht]
    \centering
    \includegraphics[width=1.\textwidth]{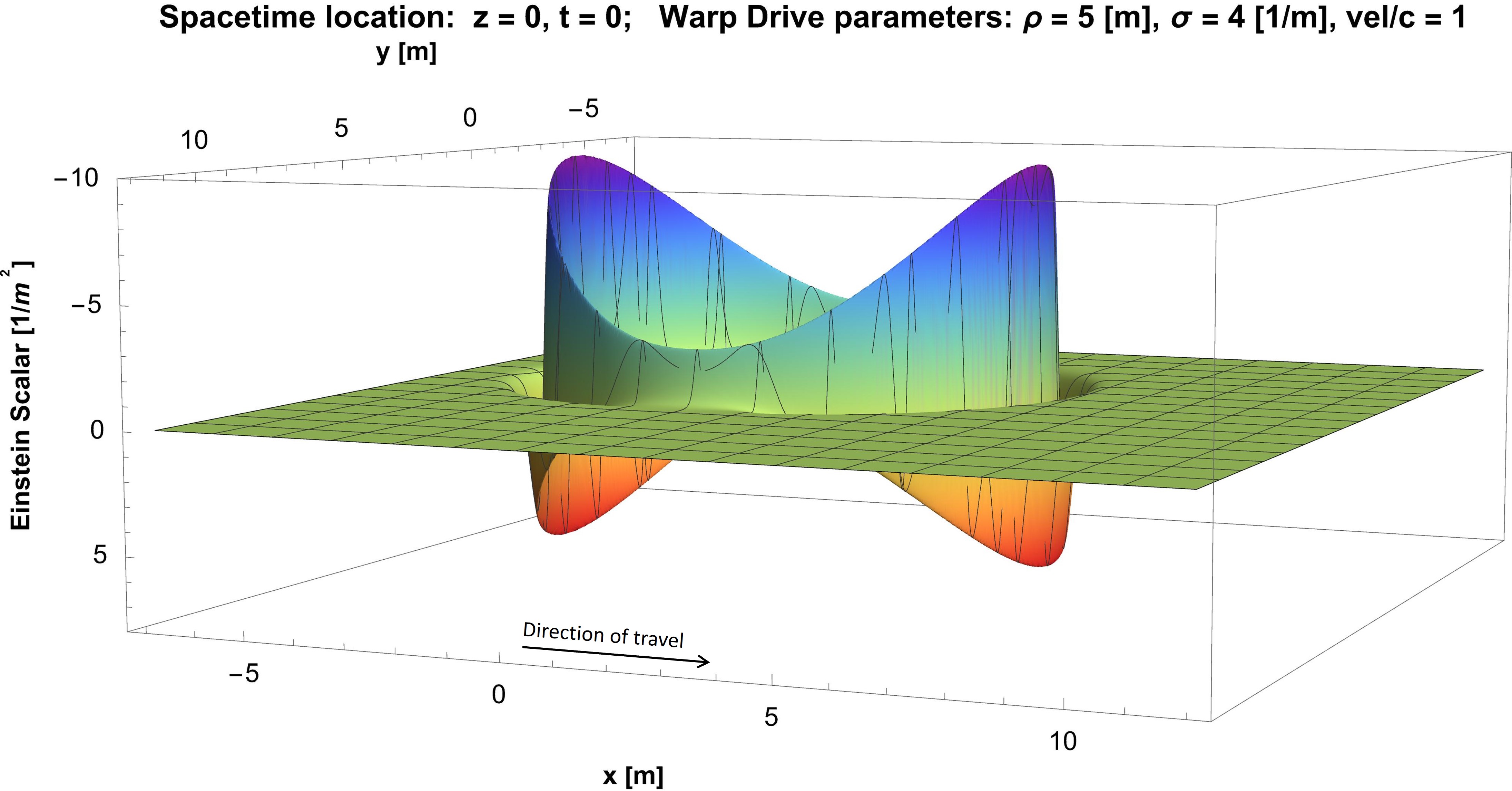}
    \caption{3-D plot of the intensity of Einstein’s curvature scalar $G= G_{\alpha}\,^{\alpha}$, of the Alcubierre warp bubble vs. the $x,y$ coordinates. The spaceship is traveling at a constant velocity (the speed of light $c$) in the positive $x$ direction, as seen by Eulerian distant external observers. A view from below, displaying negative values of Einstein's curvature scalar ($G<0$).}
    \label{Fig4_G_3D_minus}
\end{figure}

Fig.~\ref{Fig3_G_3D_plus} and Fig.~\ref{Fig4_G_3D_minus} provide three-dimensional views of the distribution of Einstein's curvature scalar $G$ in the $x,y$ plane (at $t=0$ and $z=0$) with the vertical axis representing the intensity of $G$. These plots illustrate the two spherical layers required to generate the warp bubble consistent with the distribution of $G$.\par

It is clear from Fig.~\ref{Fig3_G_3D_plus} and Fig.~\ref{Fig4_G_3D_minus} that the absolute magnitude required for negative values of Einstein's curvature scalar ($G<0$) (occurring in the inner layer and associated with positive energy density ($\rho\:c^2>0$)) is substantially larger than for positive values of Einstein's curvature scalar ($G>0$) (occurring in the outer layer and associated with positive fluid hydrostatic pressure exceeding the energy density ($3\:p>\rho\:c^2$)).

\subsection{\texorpdfstring{Quadratic Scalar Invariant of Einstein's Trace-Adjusted Curvature Tensor, \( r_1 \)}{Quadratic Scalar Invariant r1}}
% Discussion and plots related to the quadratic scalar invariant.

\begin{figure}[ht]
    \centering
    \includegraphics[width=1.\textwidth]{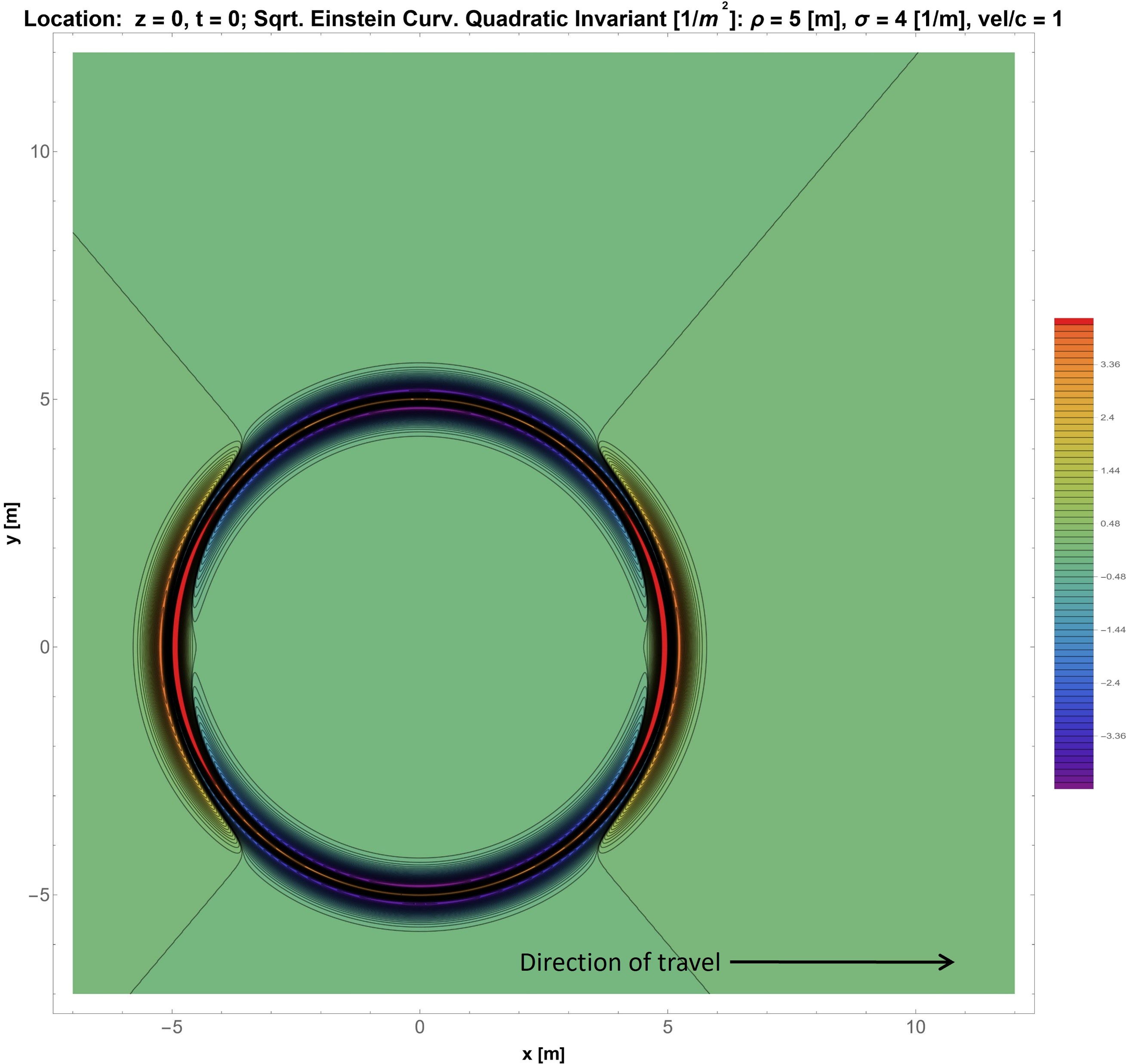}
    \caption{Contour plot in the $x,y$ plane of the signed square root $\sign{[r_{1}]}\sqrt{\abs{r_{1}}}$ of the quadratic scalar invariant
$r_{1} \equiv \widehat{R}_{\alpha}^{\:\:\:{\beta}}\widehat{R}_{\beta}^{\:\:\:{\alpha}} 
 = \widehat{G}_{\alpha}^{\:\:\:{\beta}}\widehat{G}_{\beta}^{\:\:\:{\alpha}} \; (\text{this equality valid for } n = 4)$, defined in Eq.(\ref{r1}), of the Alcubierre warp bubble. The spaceship is traveling at a constant velocity (the speed of light $c$) in the positive $x$ direction, as seen by Eulerian distant external observers.}
    \label{Fig5_G_Planform}
\end{figure}

Fig.~\ref{Fig5_G_Planform} is a contour plot (planform view) in the $x,y$ plane (at $t=0$ and $z=0$) of the quadratic scalar invariant
$r_{1} \equiv \widehat{R}_{\alpha}^{\:\:\:{\beta}}\widehat{R}_{\beta}^{\:\:\:{\alpha}} 
 = \widehat{G}_{\alpha}^{\:\:\:{\beta}}\widehat{G}_{\beta}^{\:\:\:{\alpha}} \; (\text{this equality valid for } n = 4)$, defined in Eq.(\ref{r1}), calculated based on the metric of the Alcubierre warp bubble Eq.(\ref{Alcubierre}). For direct comparison with the other plots, the square root of the absolute value of $r_{1}$ multiplied by its sign, i.e., $\sign{[r_{1}]}\sqrt{\abs{r_{1}}}$ is shown, so what is displayed has the same units $[1/m^2]$ of curvature as the previous plots for $G$ (instead of showing $r_{1}$ which has units of the square of curvature $[1/m^4]$). The Alcubierre parameters are set at $\rho=5\:[m]$, and $\sigma=4\:[1/m]$. The spaceship is traveling at a constant velocity (the speed of light $c$), in the $x$ direction, as seen by Eulerian distant external observers. \par
 
Fig.~\ref{Fig5_G_Planform} shows that the warp bubble (and hence the fuselage of the spaceship that is responsible for generating Alcubierre’s spherical warp bubble) requires up to four layers to be consistent with the distribution of $\sign{[r_{1}]}\sqrt{\abs{r_{1}}}$. As previously discussed $\sign{[r_{1}]}\sqrt{\abs{r_{1}}}$ contains information about the curvature due to simple shear deformation of the purely spatial components (shear due to planar rotation of the principal axes of stress) that is locally sourced by the fuselage as represented by the off-diagonal components of the traceless stress-energy-momentum tensor $\widehat{T}_{\alpha}^{\:\:\:{\beta}}$. An anisotropic fluid source is required to source such plane shear. The alternate sign of the four layers means that the plane shear is in opposite directions between the layers.\par

 \vspace{1em}

\begin{figure}[ht]
    \centering
    \includegraphics[width=1.\textwidth]{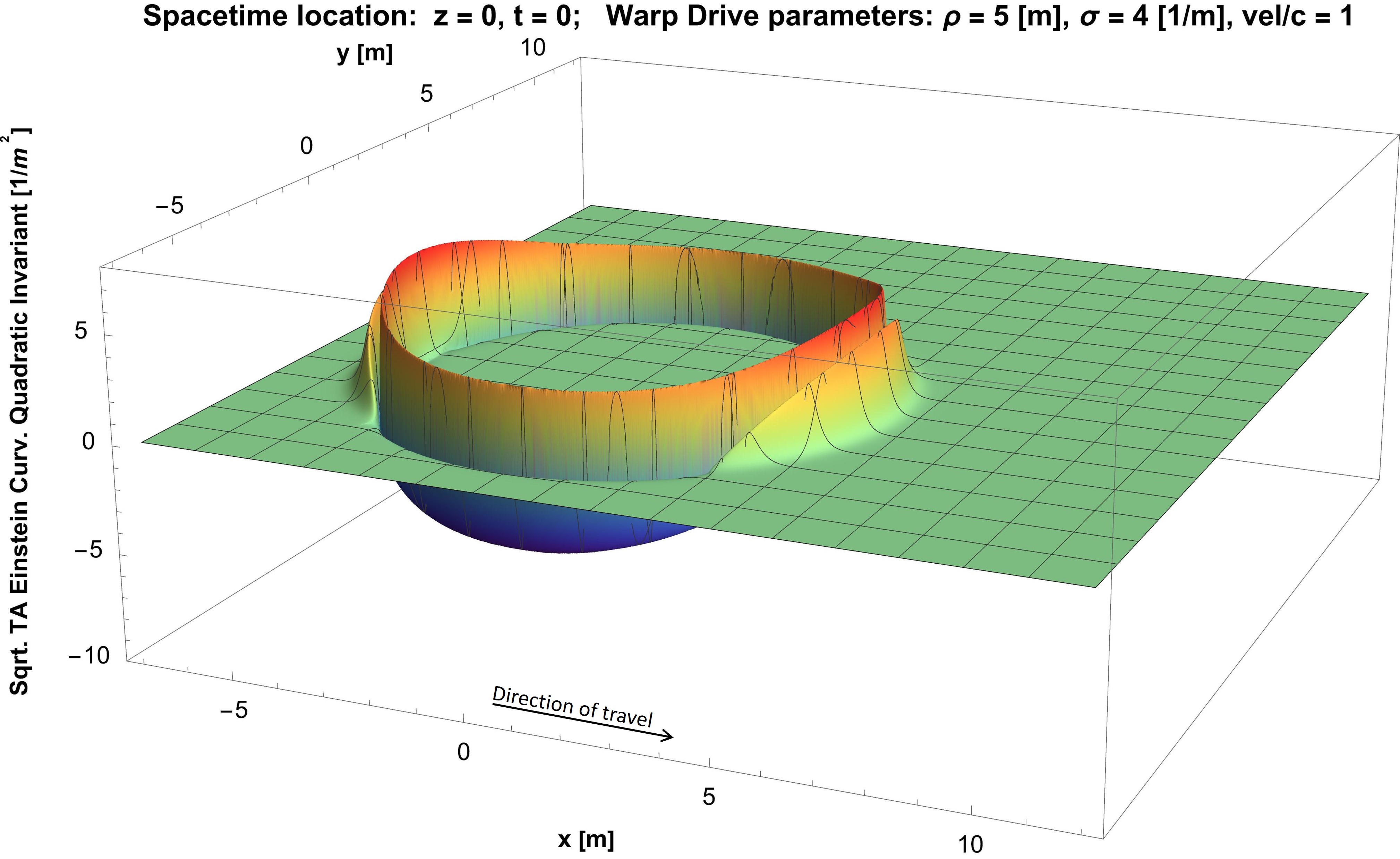}
    \caption{3-D plot of the signed square root $\sign{[r_{1}]}\sqrt{\abs{r_{1}}}$ of the quadratic scalar invariant
$r_{1} \equiv \widehat{R}_{\alpha}^{\:\:\:{\beta}}\widehat{R}_{\beta}^{\:\:\:{\alpha}} 
 = \widehat{G}_{\alpha}^{\:\:\:{\beta}}\widehat{G}_{\beta}^{\:\:\:{\alpha}} \; (\text{this equality valid for } n = 4)$, defined in Eq.(\ref{r1}), of the Alcubierre warp bubble vs. the $x,y$ coordinates. The spaceship is traveling at a constant velocity (the speed of light $c$) in the positive $x$ direction, as seen by Eulerian distant external observers.  A view from above, displaying positive values  $\sign{[r_{1}]}\sqrt{\abs{r_{1}}}>0$.}
    \label{Fig6_G_3D_plus}
\end{figure}

\begin{figure}[ht]
    \centering
    \includegraphics[width=1.\textwidth]{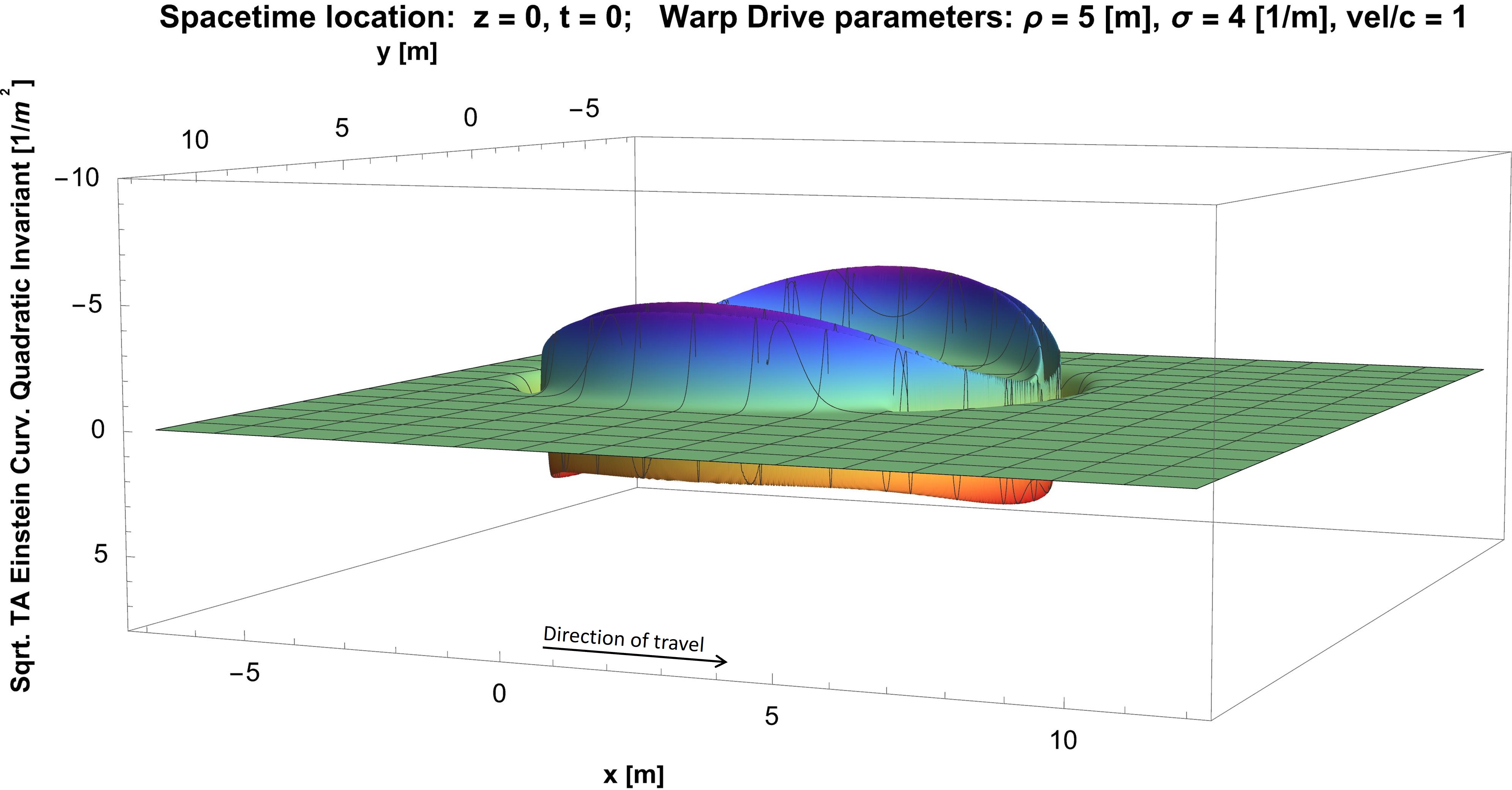}
    \caption{3-D plot of the signed square root $\sign{[r_{1}]}\sqrt{\abs{r_{1}}}$ of the quadratic scalar invariant
$r_{1} \equiv \widehat{R}_{\alpha}^{\:\:\:{\beta}}\widehat{R}_{\beta}^{\:\:\:{\alpha}} 
 = \widehat{G}_{\alpha}^{\:\:\:{\beta}}\widehat{G}_{\beta}^{\:\:\:{\alpha}} \; (\text{this equality valid for } n = 4)$, defined in Eq.(\ref{r1}), of the Alcubierre warp bubble vs. the $x,y$ coordinates. The spaceship is traveling at a constant velocity (the speed of light $c$) in the positive $x$ direction, as seen by Eulerian distant external observers. A view from below, displaying negative values $\sign{[r_{1}]}\sqrt{\abs{r_{1}}}<0$.}
    \label{Fig7_G_3D_minus}
\end{figure}

 Fig.~\ref{Fig6_G_3D_plus} and Fig.~\ref{Fig7_G_3D_minus} provide three-dimensional views of the distribution of the quadratic scalar invariant
$r_{1}$, defined in Eq.(\ref{r1}), in the $x,y$ plane (at $t=0$ and $z=0$) with the vertical axis representing the intensity of $\sign{[r_{1}]}\sqrt{\abs{r_{1}}}$. These plots illustrate the four layers required to generate the warp bubble consistent with the distribution of $\sign{[r_{1}]}\sqrt{\abs{r_{1}}}$.\par

From Fig.~\ref{Fig6_G_3D_plus} and Fig.~\ref{Fig7_G_3D_minus}, it is evident that there are two concentric layers with positive values $\sign{[r_{1}]}\sqrt{\abs{r_{1}}}>0$. The innermost layer is spherically continuous, while the outer positive layer appears only towards the front and back ends of the warp bubble in the direction of motion. Additionally, there are two concentric layers with negative values $\sign{[r_{1}]}\sqrt{\abs{r_{1}}}<0$. Both of these layers reach their maximum magnitude midstream and decrease in magnitude towards the front and back of the warp bubble in the direction of motion. 

\subsection{\texorpdfstring{Scalar Invariant of the Weyl Curvature Tensor, \( I \)}{Quadratic Scalar Invariant I}}
% Discussion and plots related to the scalar of the Weyl conformal curvature tensor.

\begin{figure}[ht]
    \centering
    \includegraphics[width=1.\textwidth]{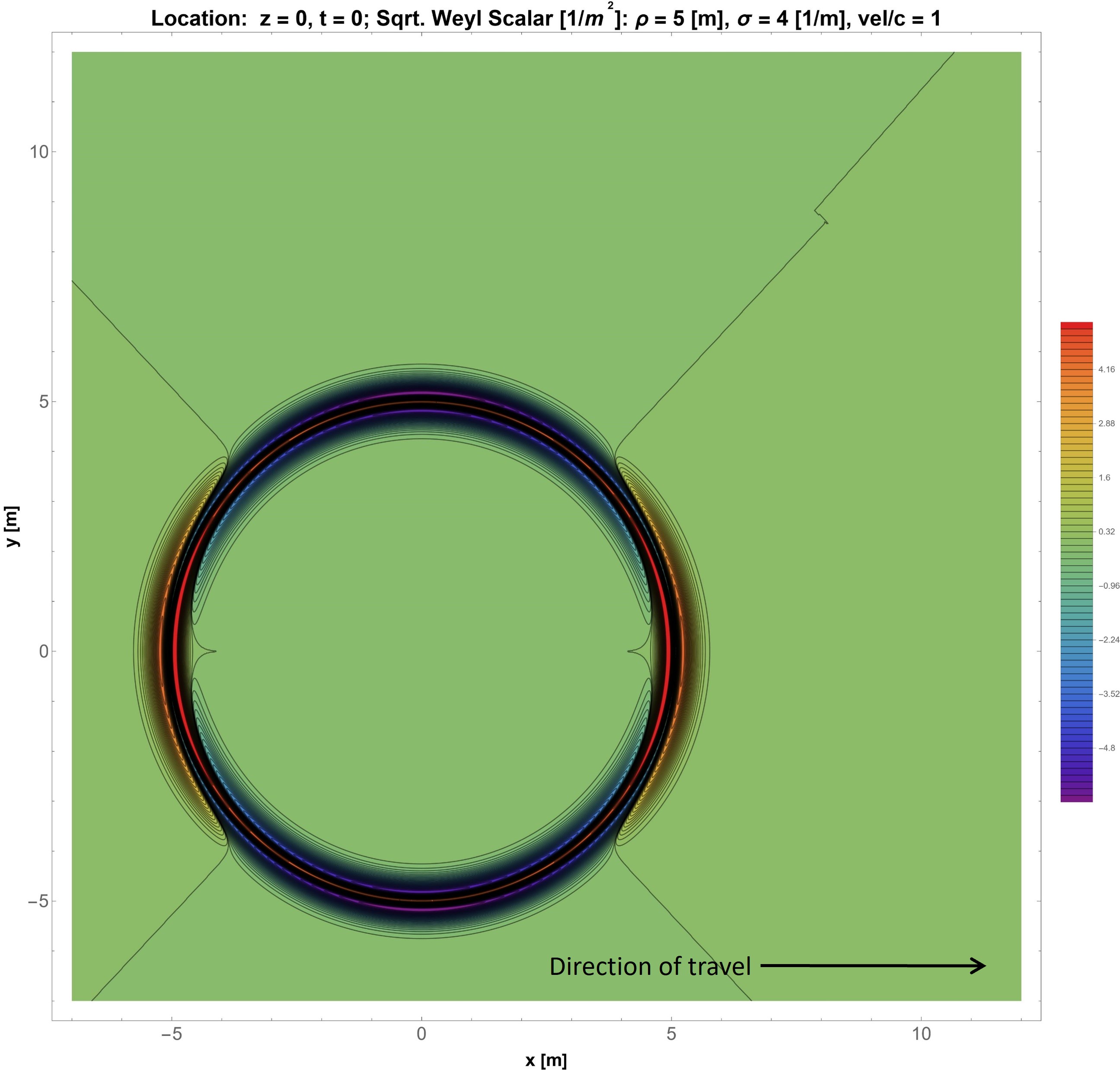}
    \caption{Contour plot in the $x,y$ plane of the signed square root $\sign{[I]}\sqrt{\abs{I}}$ of the Weyl scalar invariant
$I\equiv C_{\alpha\beta\gamma\delta} C^{\alpha\beta\gamma\delta}$, defined in Eq. ($\ref{WeylScalar}$), of the Alcubierre warp bubble. The spaceship is traveling at a constant velocity (the speed of light $c$) in the positive $x$ direction, as seen by Eulerian distant external observers.}
    \label{Fig8_G_Planform}
\end{figure}

Fig.~\ref{Fig8_G_Planform} is a contour plot (planform view) in the $x,y$ plane (at $t=0$ and $z=0$) of the Weyl scalar invariant $I\equiv C_{\alpha\beta\gamma\delta} C^{\alpha\beta\gamma\delta}$, defined in Eq. ($\ref{WeylScalar}$), using the metric Eq.(\ref{Alcubierre}) of the Alcubierre warp bubble. For direct comparison with the other plots, the square root of the absolute value of $I$ multiplied by its sign, i.e., $\sign{[I]}\sqrt{\abs{I}}$, is shown, so what is displayed has the same units $[1/m^2]$ of curvature as the previous plots (instead of showing $I$ which has units of the square of curvature $[1/m^4]$).The Alcubierre parameters are set at $\rho=5\:[m]$, and $\sigma=4\:[1/m]$. The spaceship is traveling at a constant velocity (the speed of light $c$), in the $x$ direction, as seen by Eulerian distant external observers. \par

Fig.~\ref{Fig8_G_Planform} shows that the warp bubble (and hence the fuselage of the spaceship responsible for generating Alcubierre's spherical warp bubble) requires up to four layers to be consistent with the distribution of $\sign{[I]}\sqrt{\abs{I}}$. The magnitude of the signed square root of the Weyl scalar invariant $\sign{[I]}\sqrt{\abs{I}}$ is greater than the magnitude of the signed square root of the quadratic invariant $\sign{[r_{1}]}\sqrt{\abs{r_{1}}}$, but their distribution is very similar. As previously discussed $\sign{[I]}\sqrt{\abs{I}}$ contains information about the curvature due to simple shear deformation of the purely spatial components (shear due to planar rotation of the principal axes of stress) that is non-locally sourced by the fuselage. It is non-locally sourced because it is sourced by the derivative of the Weyl tensor $C^{\alpha\beta\gamma\delta}$ that algebraically depends on the derivative of the traceless energy tensor $\widehat{T}^{\alpha \beta}$ instead of these tensors themselves being algebraically dependent. The alternating signs of the four layers mean that the plane shear is in opposite directions between the layers. An anisotropic fluid source is required to source such shear.

\vspace{1em}

\begin{figure}[ht]
    \centering
    \includegraphics[width=1.\textwidth]{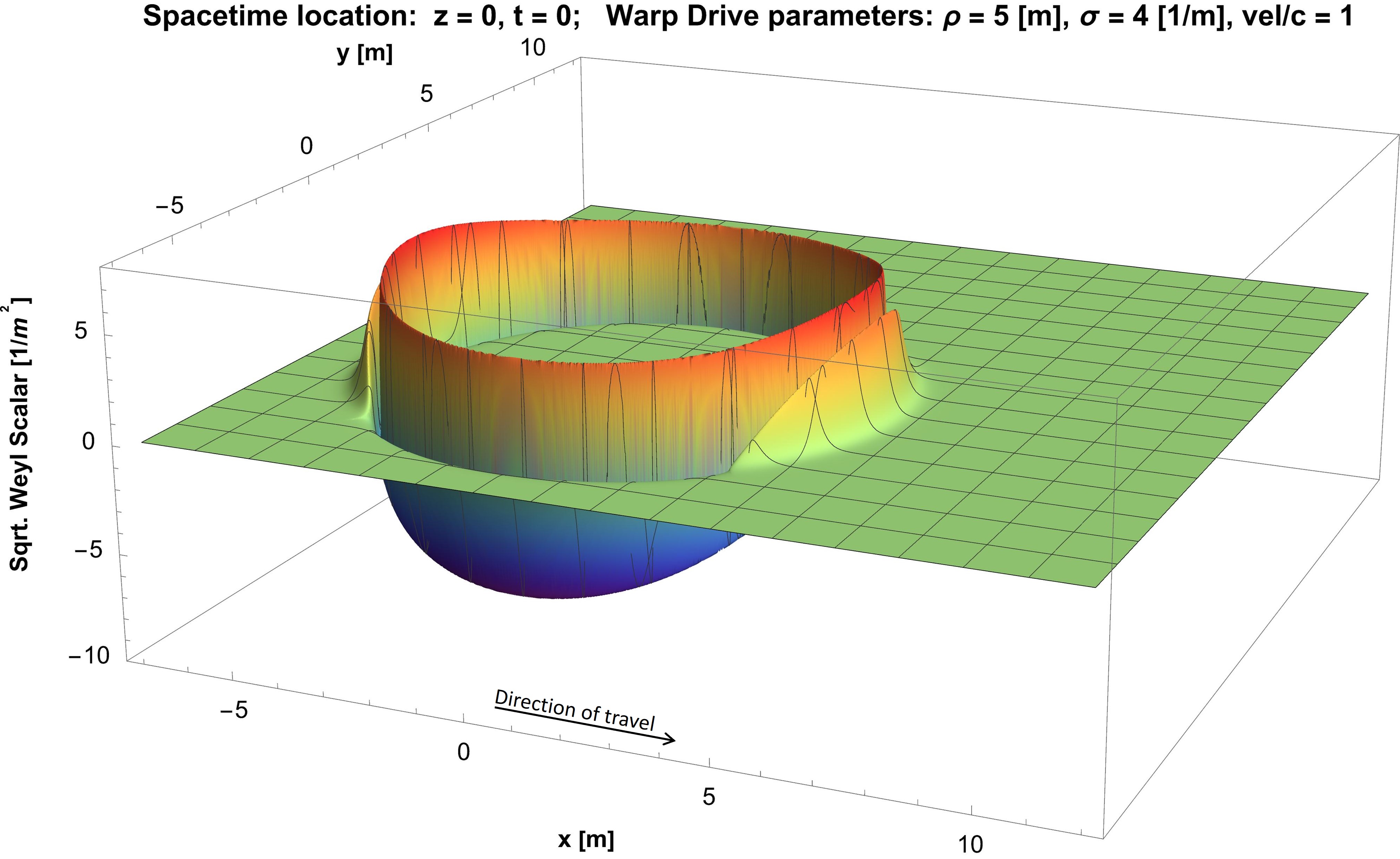}
    \caption{3-D plot of the signed square root $\sign{[I]}\sqrt{\abs{I}}$ of the Weyl scalar invariant $I\equiv C_{\alpha\beta\gamma\delta} C^{\alpha\beta\gamma\delta}$, defined in Eq. ($\ref{WeylScalar}$), of the Alcubierre warp bubble vs. the $x,y$ coordinates. The spaceship is traveling at a constant velocity (the speed of light $c$) in the positive $x$ direction, as seen by Eulerian distant external observers.  A view from above, displaying positive values  $\sign{[I]}\sqrt{\abs{I}}>0$.}
    \label{Fig9_G_3D_plus}
\end{figure}

\begin{figure}[ht]
    \centering
    \includegraphics[width=1.\textwidth]{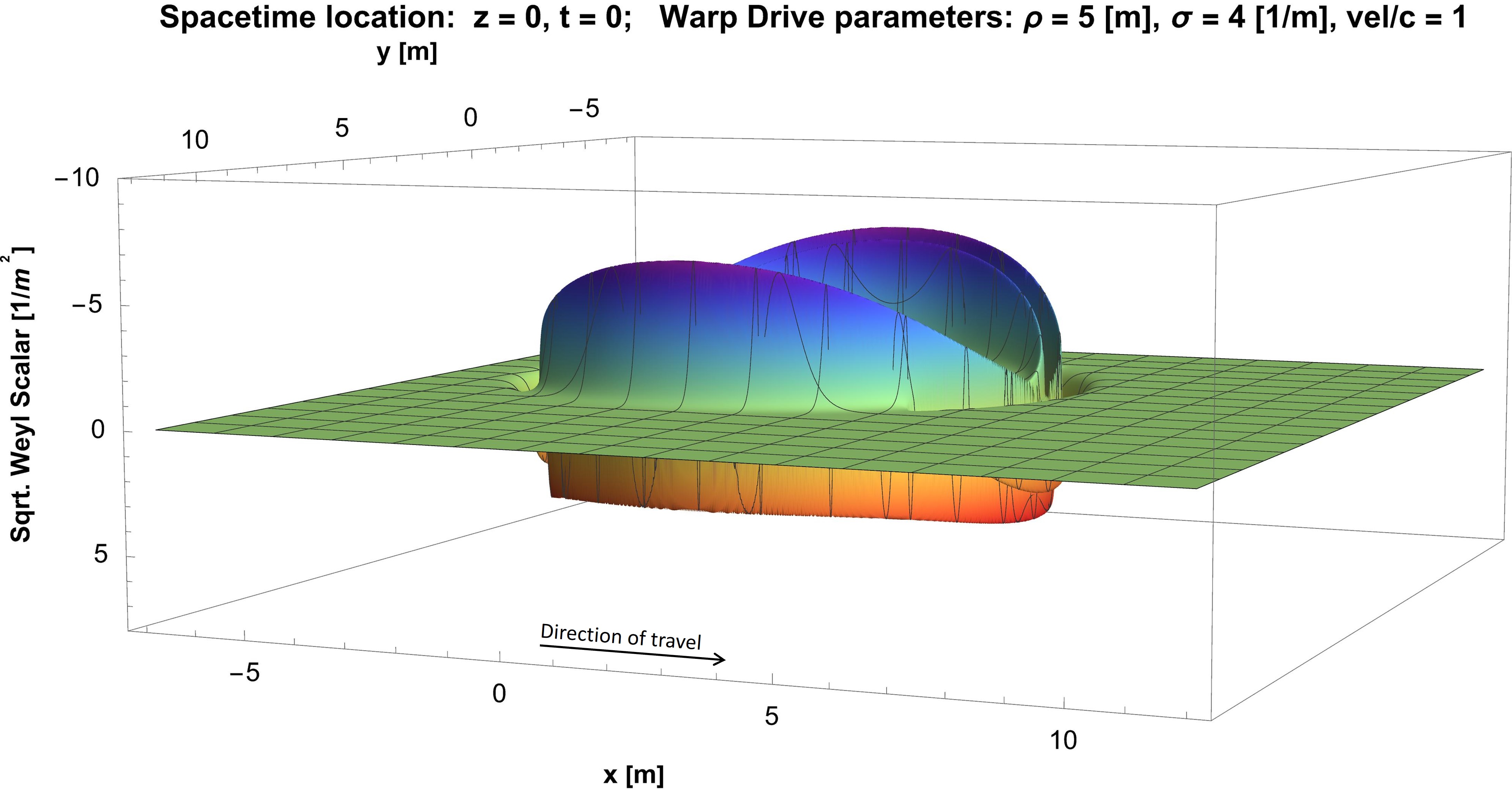}
    \caption{3-D plot of the signed square root $\sign{[I]}\sqrt{\abs{I}}$ of the Weyl scalar invariant $I\equiv C_{\alpha\beta\gamma\delta} C^{\alpha\beta\gamma\delta}$, defined in Eq. ($\ref{WeylScalar}$), of the Alcubierre warp bubble vs. the $x,y$ coordinates. The spaceship is traveling at a constant velocity (the speed of light $c$) in the positive $x$ direction, as seen by Eulerian distant external observers. A view from below, displaying negative values $\sign{[I]}\sqrt{\abs{I}}<0$.}
    \label{Fig10_G_3D_minus}
\end{figure}

Fig.~\ref{Fig9_G_3D_plus} and Fig.~\ref{Fig10_G_3D_minus} provide three-dimensional views of the distribution of the Weyl scalar invariant $I\equiv C_{\alpha\beta\gamma\delta} C^{\alpha\beta\gamma\delta}$, in the $x,y$ plane (at $t=0$ and $z=0$) with the vertical axis representing the intensity of $\sign{[I]}\sqrt{\abs{I}}$. These plots illustrate the four layers required to generate the warp bubble consistent with the distribution of $\sign{[I]}\sqrt{\abs{I}}$.\par

From Fig.~\ref{Fig9_G_3D_plus} and Fig.~\ref{Fig10_G_3D_minus}, it is evident that there are two concentric layers with positive values $\sign{[I]}\sqrt{\abs{I}}>0$. The innermost positive layer is spherically continuous, while the outer positive layer appears only in a crescent shape towards the front and back ends of the warp bubble in the direction of motion. Additionally, there are two crescent-shaped concentric layers with negative values $\sign{[I]}\sqrt{\abs{I}}<0$. Both of these layers reach their maximum magnitude midstream and decrease in magnitude towards the front and back of the warp bubble in the direction of motion. The previous planform view indicates that the two crescent-shaped concentric layers with negative values of the signed square root of the Weyl scalar $\sign{[I]}\sqrt{\abs{I}}$ have zero magnitude along the $x$ axis of motion. The innermost \textit{contour} of the signed square root of the Weyl scalar $\sign{[I]}\sqrt{\abs{I}}$ at these locations exhibits a reentrant singularity.

\subsection{\texorpdfstring{Cubic Scalar Invariant of Einstein's Trace-Adjusted Curvature Tensor, \( r_2 \)}{Cubic Scalar Invariant r2}}
% Discussion and plots related to the cubic scalar invariant.

\begin{figure}[ht]
    \centering
    \includegraphics[width=1.\textwidth]{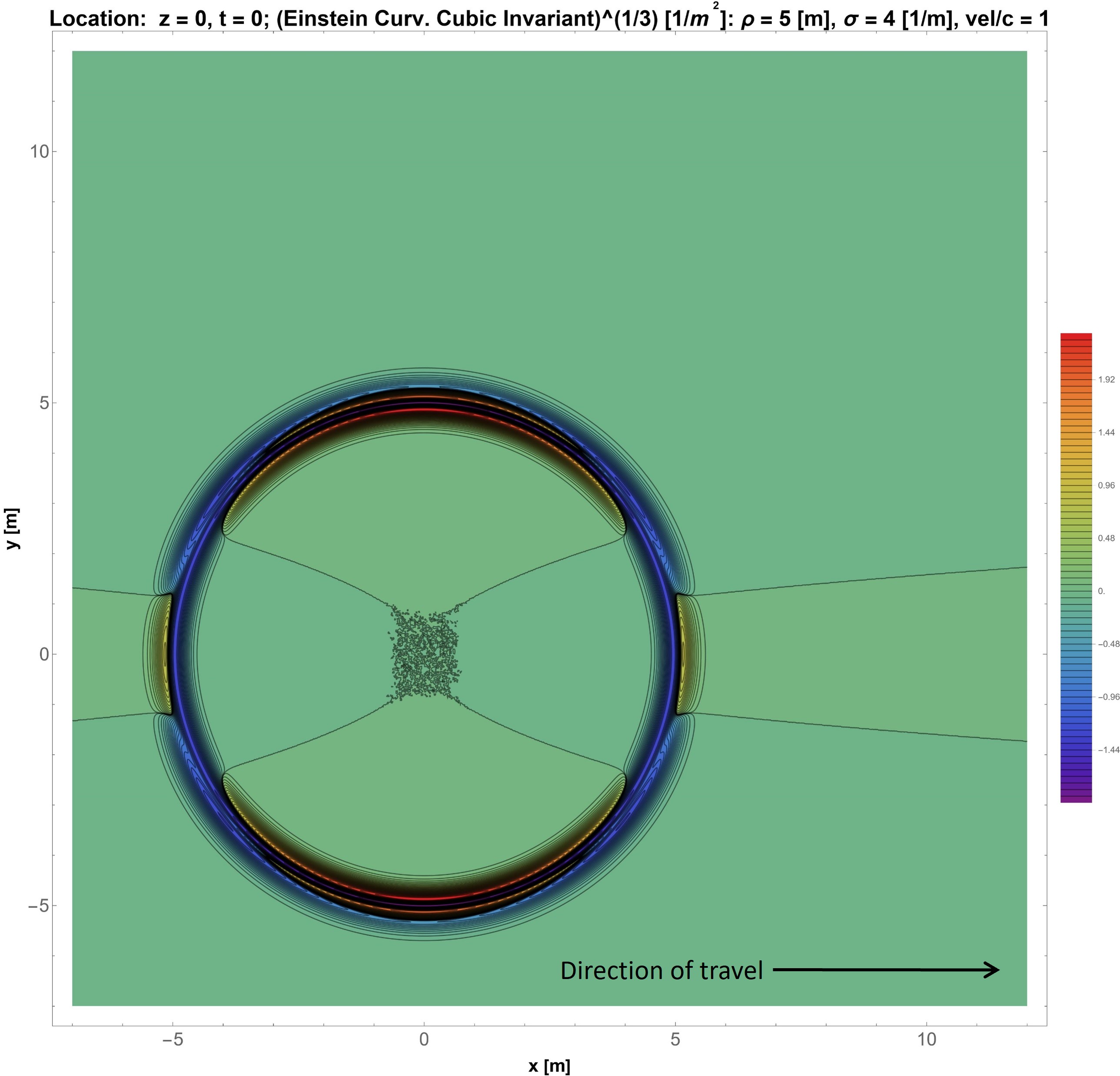}
    \caption{Contour plot in the $x,y$ plane of the signed cubic root $\sign{[r_{2}]}\sqrt[3]{\abs{r_{2}}}$ of the cubic scalar invariant
$r_{2} \equiv \widehat{R}_{\alpha}^{\:\:\:{\beta}}\widehat{R}_{\beta}^{\:\:\:{\gamma}} \widehat{R}_{\gamma}^{\:\:\:{\alpha}} = \widehat{G}_{\alpha}^{\:\:\:{\beta}}\widehat{G}_{\beta}^{\:\:\:{\gamma}} \widehat{G}_{\gamma}^{\:\:\:{\alpha}}\: \text{(this equality valid for } n = 4)$, defined in Eq.(\ref{r2}), of the Alcubierre warp bubble. The spaceship is traveling at a constant velocity (the speed of light $c$) in the positive $x$ direction, as seen by Eulerian distant external observers.}
    \label{Fig11_G_Planform}
\end{figure}

Fig.~\ref{Fig11_G_Planform} is a contour plot (planform view) in the $x,y$ plane (at $t=0$ and $z=0$) of the cubic scalar invariant
$r_{2} \equiv \widehat{R}_{\alpha}^{\:\:\:{\beta}}\widehat{R}_{\beta}^{\:\:\:{\gamma}} \widehat{R}_{\gamma}^{\:\:\:{\alpha}} = \widehat{G}_{\alpha}^{\:\:\:{\beta}}\widehat{G}_{\beta}^{\:\:\:{\gamma}} \widehat{G}_{\gamma}^{\:\:\:{\alpha}}\; \text{(this equality valid for } n = 4)$, defined in Eq.(\ref{r2}), calculated based on the metric of the Alcubierre warp bubble Eq.(\ref{Alcubierre}). For direct comparison with the other plots, the cubic root of the absolute value of $r_{2}$ multiplied by its sign, i.e., $\sign{[r_{2}]}\sqrt[3]{\abs{r_{2}}}$ is shown, so what is displayed has the same units $[1/m^2]$ of curvature as the previous plots (instead of showing $r_{2}$ which has units of the cube of curvature $[1/m^6]$). The Alcubierre parameters are set at $\rho=5\:[m]$, and $\sigma=4\:[1/m]$. The spaceship is traveling at a constant velocity (the speed of light $c$), in the $x$ direction, as seen by Eulerian distant external observers.\par 

Fig.~\ref{Fig11_G_Planform} shows that the warp bubble (and hence the fuselage of the spaceship responsible for generating Alcubierre's spherical warp bubble) requires up to four layers to be consistent with the distribution of $\sign{[r_{2}]}\sqrt[3]{\abs{r_{2}}}$. The magnitude of the cubic scalar invariant $\sign{[r_{2}]}\sqrt[3]{\abs{r_{2}}}$ is significantly smaller than the magnitude of the quadratic invariant $\sign{[r_{1}]}\sqrt{\abs{r_{1}}}$. Their distribution also differs considerably. As previously discussed $\sign{[r_{2}]}\sqrt[3]{\abs{r_{2}}}$ contains information about the curvature due to three-dimensional shear deformation of the purely spatial components (shear due to three-dimensional rotation of the principal axes of stress) that is locally sourced by the fuselage as represented by the off-diagonal components of the traceless stress-energy-momentum tensor $\widehat{T}_{\alpha}^{\:\:\:{\beta}}$. An anisotropic fluid source is required to source such three-dimensional shear. The alternating signs of the four layers mean that the three-dimensional shear is in opposite directions between the layers. The fractal-looking small contours at the center of the warp bubble are due to numerical round-off resulting from the very small magnitude of $\sign{[r_{2}]}\sqrt[3]{\abs{r_{2}}}$ at that location and should be considered a numerical artifact. Analytical inspection of the equations involved in the calculation of contours revealed (performing the Limit operation) that the interior region of the warp bubble is indeed perfectly smooth Minkowski flat and there is no fractality to it. It is noteworthy that the cubic invariant $r_{2}$ does not vanish in Alcubierre’s spacetime (as asserted in \cite{Mattingly}) and that Alcubierre’s spacetime is not a Class $B$ warped product spacetime (as claimed in \cite{Mattingly} without proof), as discussed in Section~\ref{sec:analysis_alcubierre}.\par

\begin{figure}[ht]
    \centering
    \includegraphics[width=1.\textwidth]{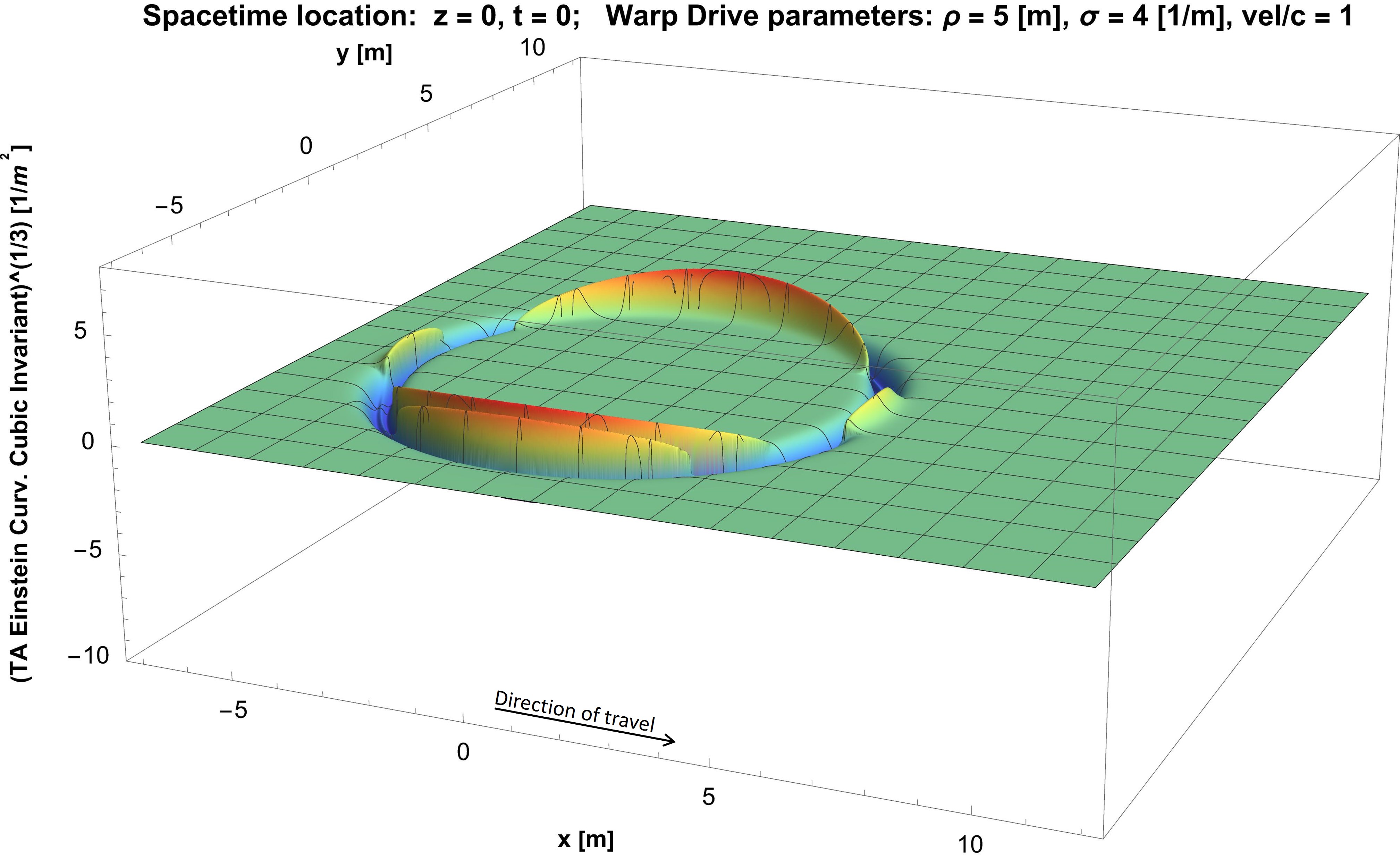}
    \caption{3-D plot of the signed cubic root $\sign{[r_{2}]}\sqrt[3]{\abs{r_{2}}}$ of the cubic scalar invariant
$r_{2} \equiv \widehat{R}_{\alpha}^{\:\:\:{\beta}}\widehat{R}_{\beta}^{\:\:\:{\gamma}} \widehat{R}_{\gamma}^{\:\:\:{\alpha}} = \widehat{G}_{\alpha}^{\:\:\:{\beta}}\widehat{G}_{\beta}^{\:\:\:{\gamma}} \widehat{G}_{\gamma}^{\:\:\:{\alpha}}\: \text{(this equality valid for } n = 4)$, defined in Eq.(\ref{r2}), of the Alcubierre warp bubble vs. the $x,y$ coordinates. The spaceship is traveling at a constant velocity (the speed of light $c$) in the positive $x$ direction, as seen by Eulerian distant external observers.  A view from above, displaying positive values  $\sign{[r_{2}]}\sqrt[3]{\abs{r_{2}}}>0$.}
    \label{Fig12_G_3D_plus}
\end{figure}

\begin{figure}[ht]
    \centering
    \includegraphics[width=1.\textwidth]{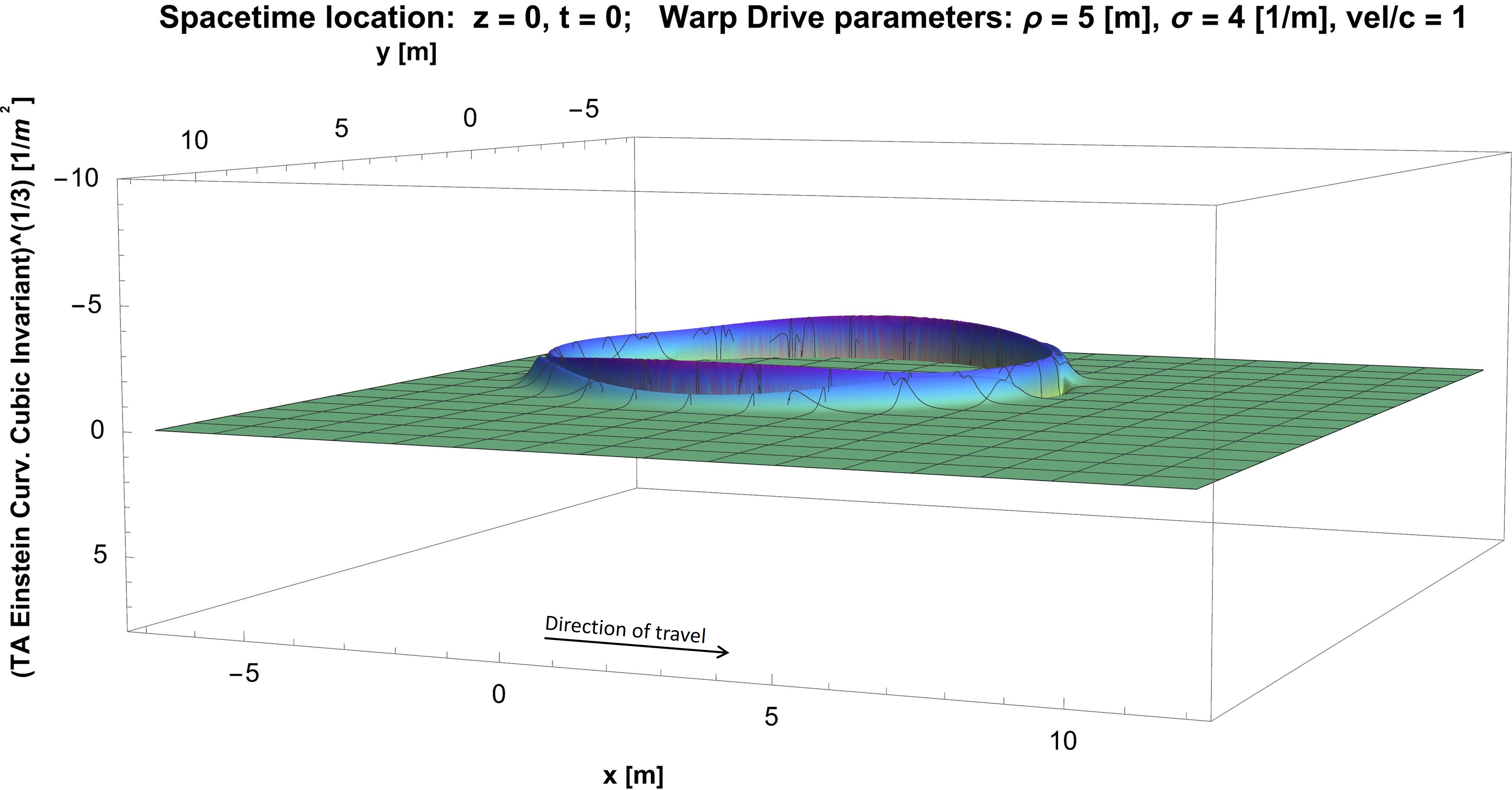}
    \caption{3-D plot of the signed cubic root $\sign{[r_{2}]}\sqrt[3]{\abs{r_{2}}}$ of the cubic scalar invariant
$r_{2} \equiv \widehat{R}_{\alpha}^{\:\:\:{\beta}}\widehat{R}_{\beta}^{\:\:\:{\gamma}} \widehat{R}_{\gamma}^{\:\:\:{\alpha}} = \widehat{G}_{\alpha}^{\:\:\:{\beta}}\widehat{G}_{\beta}^{\:\:\:{\gamma}} \widehat{G}_{\gamma}^{\:\:\:{\alpha}}\: \text{(this equality valid for } n = 4)$, defined in Eq.(\ref{r2}), of the Alcubierre warp bubble vs. the $x,y$ coordinates. The spaceship is traveling at a constant velocity (the speed of light $c$) in the positive $x$ direction, as seen by Eulerian distant external observers. A view from below, displaying negative values $\sign{[r_{2}]}\sqrt[3]{\abs{r_{2}}}<0$.}
    \label{Fig13_G_3D_minus}
\end{figure}

Fig.~\ref{Fig12_G_3D_plus} and Fig.~\ref{Fig13_G_3D_minus} provide three-dimensional views of the distribution of the cubic scalar invariant
$r_{2}$, defined in Eq.(\ref{r2}), in the $x,y$ plane (at $t=0$ and $z=0$) with the vertical axis representing the intensity of $\sign{[r_{2}]}\sqrt{\abs{r_{2}}}$. 

These plots also illustrate the four layers required to generate the warp bubble consistent with the distribution of $\sign{[r_{2}]}\sqrt[3]{\abs{r_{2}}}$.  From Fig.~\ref{Fig12_G_3D_plus} and Fig.~\ref{Fig13_G_3D_minus}, it is evident that there are two crescent-shaped concentric layers with positive values $\sign{[r_{2}]}\sqrt[3]{\abs{r_{2}}}>0$, reaching highest value midstream between the front and back of the warp bubble and diminishing to zero and hence vanishing at both ends in the $x$ direction of motion. The outermost crescent-shaped positive layer re-emerges again at the front and back of the warp bubble. Additionally, there are two concentric layers with negative values $\sign{[r_{2}]}\sqrt[3]{\abs{r_{2}}}<0$.  The innermost layer of negative value $\sign{[r_{2}]}\sqrt[3]{\abs{r_{2}}}<0$ is spherically continuous. The outermost layer with negative value $\sign{[r_{2}]}\sqrt[3]{\abs{r_{2}}}<0$  is crescent shaped and has zero value at the front and back of the warp bubble in the $x$ direction of motion.

\section{Conclusions}

Upon analyzing the Alcubierre warp-drive spacetime and investigating scalar curvature invariants, we have arrived at several key findings:

\begin{enumerate}
    \item The distribution (based on the Alcubierre metric and form function defined in Eq.(\ref{Alcubierre},\ref{AlcubierreForm})) of the quadratic scalar invariant
$r_{1} \equiv \widehat{R}_{\alpha}^{\:\:\:{\beta}}\widehat{R}_{\beta}^{\:\:\:{\alpha}} 
 = \widehat{G}_{\alpha}^{\:\:\:{\beta}}\widehat{G}_{\beta}^{\:\:\:{\alpha}} \; (\text{this equality valid for } n = 4)$, defined in Eq.(\ref{r1}), of the cubic scalar invariant
$r_{2} \equiv \widehat{R}_{\alpha}^{\:\:\:{\beta}}\widehat{R}_{\beta}^{\:\:\:{\gamma}} \widehat{R}_{\gamma}^{\:\:\:{\alpha}} = \widehat{G}_{\alpha}^{\:\:\:{\beta}}\widehat{G}_{\beta}^{\:\:\:{\gamma}} \widehat{G}_{\gamma}^{\:\:\:{\alpha}}\: \text{(this equality valid for } n = 4)$, defined in Eq.(\ref{r2}), and of the Weyl scalar invariant $I\equiv C_{\alpha\beta\gamma\delta} C^{\alpha\beta\gamma\delta}$, defined in Eq. ($\ref{WeylScalar}$), necessitate the presence of four concentric layers of an anisotropic fluid to be consistent with the distribution of the Alcubierre warp bubble.
    
    \item The distribution of the Einstein curvature scalar $G=G_{\alpha}\,^{\alpha}$ necessitates the presence of two concentric layers in an anisotropic fluid to be consistent with the distribution of the Alcubierre warp bubble.
    
    \item The layers exhibit distinct distributions, with some being spherically continuous while others appear in crescent shapes; several of them vanish in the direction of motion.
    
    \item The magnitude and distribution of these invariants provide physical insights into the nature of the curvature due to volume change and of the curvature due to planar and three-dimensional shear deformation within the warp bubble. Additionally, they highlight the need for a stress-energy-momentum tensor with diagonal and off-diagonal components characteristic of an anisotropic fluid.

\item The spacetime invariant curvature of the Alcubierre warp-drive traveling at the speed of light, when accurately represented, is comparable to the invariant curvature at the event horizon of a Schwarzschild black hole with the mass equivalent of the planet Saturn.

\item While the study by Mattingly et al. \cite{Mattingly} has advanced the understanding of the curvature invariants of the Alcubierre warp-drive spacetime, it exhibits inconsistencies in both units and graphical data representation. Their decision to truncate the curvature invariants by an extensive 8 to 16 orders of magnitude seems both arbitrary and unjustified. These discrepancies, coupled with potential misrepresentations stemming from the default plotting range behavior of Wolfram \textit{Mathematica\textsuperscript{\textregistered}} \cite{Mathematica}, highlight the critical importance of precise representation of spacetime curvature changes in any discourse on the Alcubierre warp-drive.

\item The Alcubierre metric, based on the 3+1 formalism, ensures a specific causal structure locally but doesn’t guarantee global hyperbolicity. The Alcubierre metric showcases warping of spacetime but does not meet the specific criteria set for ``Class $B$'' warped product spacetimes.  The form function in the Alcubierre metric intertwines all spacetime coordinates, preventing a straightforward decomposition into two distinct Lorentzian and Riemannian 2-dimensional spaces.

\end{enumerate}

Furthermore, this study delivers a precise visualization of scalar invariants within the Alcubierre warp bubble and emphasizes their fundamental role in elucidating spacetime configurations in general relativity.

\section*{Declarations}

\begin{itemize}
\item Funding: No funding was received for conducting this study.
\item Conflict of interest/Competing interests: The author has no competing interests to declare that are relevant to the content of this article.
\end{itemize}

\noindent


\begin{appendices}

\section{Curvature Invariants of the Alcubierre metric}\label{secA1}

\vspace{1cm}

\begin{equation}
f_r =\frac{\tanh{[\sigma (\sqrt{(x-x_{0}(t))^2+y^2+z^2}+\rho)}]-\tanh{[\sigma (\sqrt{(x-x_{0}(t))^2+y^2+z^2}-\rho)}]}{2\tanh{[\sigma \rho]}}
\label{AlcubierreForm2}
\end{equation}\par

\vspace{1cm}

\begin{equation}
v_s =\frac{\partial x_{0}(t)}{\partial t}
\label{Velocity}
\end{equation}\par

\vspace{1cm}

\begin{equation}
\begin{aligned}
G &=-\frac{1}{2} v_s \left(v_s \left(\frac{\partial f_r}{\partial z}^2+\frac{\partial f_r}{\partial y}^2+4 \left(\frac{\partial f_r}{\partial x}^2+f_r \frac{\partial^2 f_r}{\partial x^2}\right)\right)+4 \frac{\partial^2 f_r}{\partial t \partial x}\right)\\
\end{aligned}
\label{Ginv}
\end{equation}\par

\vspace{1cm}

\begin{equation}
\begin{aligned}
r_1 &=\frac{1}{16} v_s^2 \biggl(11 v_s^2 \frac{\partial f_r}{\partial z}^4+11 v_s^2 \frac{\partial f_r}{\partial y}^4+8 (2 v_s^2 \frac{\partial f_r}{\partial x}^4+v_s^2 f_r^2 (\frac{\partial^2 f_r}{\partial x \partial z}^2+\frac{\partial^2 f_r}{\partial x \partial y}^2+
2 \frac{\partial^2 f_r}{\partial x^2}^2)\\
&\quad 
+4 v_s \frac{\partial f_r}{\partial x}^2 (v_s f_r \frac{\partial^2 f_r}{\partial x^2}+\frac{\partial^2 f_r}{\partial t \partial x})+2 v_s f_r (\frac{\partial^2 f_r}{\partial x \partial z} \frac{\partial^2 f_r}{\partial t \partial z}+\frac{\partial^2 f_r}{\partial x \partial y} \frac{\partial^2 f_r}{\partial t \partial y}+2 \frac{\partial^2 f_r}{\partial x^2} \frac{\partial^2 f_r}{\partial t \partial x})\\
&\quad 
-\frac{\partial^2 f_r}{\partial x \partial z}^2-\frac{\partial^2 f_r}{\partial x \partial y}^2+\frac{\partial^2 f_r}{\partial t \partial z}^2+\frac{\partial^2 f_r}{\partial t \partial y}^2+2 \frac{\partial^2 f_r}{\partial t \partial x}^2)+\\
&\quad 
2 v_s \frac{\partial f_r}{\partial z}^2 (11 v_s \frac{\partial f_r}{\partial y}^2-4 (-3 v_s \frac{\partial f_r}{\partial x}^2+v_s f_r \frac{\partial^2 f_r}{\partial x^2}+\frac{\partial^2 f_r}{\partial t \partial x}))+\\
&\quad 
32 v_s \frac{\partial f_r}{\partial x} \frac{\partial f_r}{\partial z} (v_s f_r \frac{\partial^2 f_r}{\partial x \partial z}+\frac{\partial^2 f_r}{\partial t \partial z})+
32 v_s \frac{\partial f_r}{\partial y} \frac{\partial f_r}{\partial x} (v_s f_r \frac{\partial^2 f_r}{\partial x \partial y}+\frac{\partial^2 f_r}{\partial t \partial y})-\\
&\quad 
8 v_s \frac{\partial f_r}{\partial y}^2 (-3 v_s \frac{\partial f_r}{\partial x}^2+v_s f_r \frac{\partial^2 f_r}{\partial x^2}+\frac{\partial^2 f_r}{\partial t \partial x})-8 (\frac{\partial^2 f_r}{\partial z^2}+\frac{\partial^2 f_r}{\partial y^2})^2\biggl)
\end{aligned}
\label{r1long}
\end{equation}

\vspace{1cm}

\begin{equation}
\begin{aligned}
I &=\frac{1}{3} v_s^2 \biggl(4 v_s^2  \frac{\partial f_r}{\partial z}^4+4 v_s^2  \frac{\partial f_r}{\partial y}^4+4 v_s^2  \frac{\partial f_r}{\partial x}^4+3 v_s^2 f_r^2 \frac{\partial^2 f_r}{\partial x \partial z}^2+3 v_s^2 f_r^2 \frac{\partial^2 f_r}{\partial x \partial y}^2+\\
&\quad 
4 v_s^2 f_r^2 \frac{\partial^2 f_r}{\partial x^2}^2+8 v_s^2 f_r  \frac{\partial f_r}{\partial x}^2 \frac{\partial^2 f_r}{\partial x^2}+4 v_s  \frac{\partial f_r}{\partial z}^2 (2 v_s ( \frac{\partial f_r}{\partial y}^2+ \frac{\partial f_r}{\partial x}^2)-v_s f_r \frac{\partial^2 f_r}{\partial x^2}-\frac{\partial^2 f_r}{\partial t \partial x})\\
&\quad 
+12 v_s  \frac{\partial f_r}{\partial x}  \frac{\partial f_r}{\partial z} (v_s f_r \frac{\partial^2 f_r}{\partial x \partial z}+\frac{\partial^2 f_r}{\partial t \partial z})+6 v_s f_r \frac{\partial^2 f_r}{\partial x \partial z} \frac{\partial^2 f_r}{\partial t \partial z}+6 v_s f_r \frac{\partial^2 f_r}{\partial x \partial y} \frac{\partial^2 f_r}{\partial t \partial y}+\\
&\quad 
12 v_s  \frac{\partial f_r}{\partial y}  \frac{\partial f_r}{\partial x} (v_s f_r \frac{\partial^2 f_r}{\partial x \partial y}+\frac{\partial^2 f_r}{\partial t \partial y})+4 v_s  \frac{\partial f_r}{\partial y}^2 (2 v_s  \frac{\partial f_r}{\partial x}^2-v_s f_r \frac{\partial^2 f_r}{\partial x^2}-\frac{\partial^2 f_r}{\partial t \partial x})+\\
&\quad 
8 v_s ( \frac{\partial f_r}{\partial x}^2+f_r \frac{\partial^2 f_r}{\partial x^2}) \frac{\partial^2 f_r}{\partial t \partial x}+3 \frac{\partial^2 f_r}{\partial t \partial z}^2+3 \frac{\partial^2 f_r}{\partial t \partial y}^2+4 \frac{\partial^2 f_r}{\partial t \partial x}^2-\\
&\quad 
3 (4 \frac{\partial^2 f_r}{\partial y \partial z}^2+(\frac{\partial^2 f_r}{\partial z^2}-\frac{\partial^2 f_r}{\partial y^2})^2)-3 (\frac{\partial^2 f_r}{\partial x \partial z}^2+\frac{\partial^2 f_r}{\partial x \partial y}^2)\biggl)
\end{aligned}
\label{weyllong}
\end{equation}

\vspace{1cm}

\begin{equation}
\begin{aligned}
r_2 &=-\frac{3}{64}  v_s^3  \biggl(3 v_s^3  \frac{\partial f_r}{\partial z}^6+v_s^2 (9 v_s  \frac{\partial f_r}{\partial y}^2+4 v_s  \frac{\partial f_r}{\partial x}^2-12 (v_s f_r \frac{\partial^2 f_r}{\partial x^2}+\frac{\partial^2 f_r}{\partial t \partial x}))  \frac{\partial f_r}{\partial z}^4+\\
&\quad 
16 v_s^2  \frac{\partial f_r}{\partial x} (v_s f_r \frac{\partial^2 f_r}{\partial x \partial z}+\frac{\partial^2 f_r}{\partial t \partial z})  \frac{\partial f_r}{\partial z}^3+v_s (9 v_s^2  \frac{\partial f_r}{\partial y}^4+8 v_s (v_s  \frac{\partial f_r}{\partial x}^2-\\
&\quad 
3 (v_s f_r \frac{\partial^2 f_r}{\partial x^2}+\frac{\partial^2 f_r}{\partial t \partial x}))  \frac{\partial f_r}{\partial y}^2+16 v_s  \frac{\partial f_r}{\partial x} (v_s f_r \frac{\partial^2 f_r}{\partial x \partial y}+\frac{\partial^2 f_r}{\partial t \partial y})  \frac{\partial f_r}{\partial y}-\\
&\quad 
4 (\frac{\partial^2 f_r}{\partial z^2}+\frac{\partial^2 f_r}{\partial y^2})^2+4 ((v_s^2 f_r^2+3) \frac{\partial^2 f_r}{\partial x \partial z}^2+2 v_s f_r \frac{\partial^2 f_r}{\partial t \partial z} \frac{\partial^2 f_r}{\partial x \partial z}+\frac{\partial^2 f_r}{\partial x \partial y}^2+\frac{\partial^2 f_r}{\partial t \partial z}^2+\\
&\quad 
3 (v_s f_r \frac{\partial^2 f_r}{\partial x \partial y}+\frac{\partial^2 f_r}{\partial t \partial y})^2))  \frac{\partial f_r}{\partial z}^2+16 v_s (-v_s^2  \frac{\partial f_r}{\partial y} \frac{\partial^2 f_r}{\partial x \partial z} \frac{\partial^2 f_r}{\partial x \partial y} f_r^2+\\
&\quad 
v_s  \frac{\partial f_r}{\partial y} (\frac{\partial^2 f_r}{\partial x \partial z} (v_s  \frac{\partial f_r}{\partial y}  \frac{\partial f_r}{\partial x}-\frac{\partial^2 f_r}{\partial t \partial y})-\frac{\partial^2 f_r}{\partial x \partial y} \frac{\partial^2 f_r}{\partial t \partial z}) f_r-2 \frac{\partial^2 f_r}{\partial z^2}  \frac{\partial f_r}{\partial x} \frac{\partial^2 f_r}{\partial x \partial z}-\\
&\quad 
2 \frac{\partial^2 f_r}{\partial y^2}  \frac{\partial f_r}{\partial x} \frac{\partial^2 f_r}{\partial x \partial z}+ \frac{\partial f_r}{\partial y} \frac{\partial^2 f_r}{\partial x \partial z} \frac{\partial^2 f_r}{\partial x \partial y}+v_s  \frac{\partial f_r}{\partial y}^2  \frac{\partial f_r}{\partial x} \frac{\partial^2 f_r}{\partial t \partial z}- \frac{\partial f_r}{\partial y} \frac{\partial^2 f_r}{\partial t \partial z} \frac{\partial^2 f_r}{\partial t \partial y})  \frac{\partial f_r}{\partial z}+\\
&\quad 
3 v_s^3  \frac{\partial f_r}{\partial y}^6-4 v_s  \frac{\partial f_r}{\partial y}^2 \frac{\partial^2 f_r}{\partial y^2}^2+4 v_s^3  \frac{\partial f_r}{\partial y}^4  \frac{\partial f_r}{\partial x}^2+16 v_s \frac{\partial^2 f_r}{\partial y^2}^2  \frac{\partial f_r}{\partial x}^2+\\
&\quad 
12 v_s^3 f_r^2  \frac{\partial f_r}{\partial y}^2 \frac{\partial^2 f_r}{\partial x \partial z}^2+4 v_s  \frac{\partial f_r}{\partial y}^2 \frac{\partial^2 f_r}{\partial x \partial z}^2-16 v_s f_r \frac{\partial^2 f_r}{\partial y^2} \frac{\partial^2 f_r}{\partial x \partial z}^2+4 v_s^3 f_r^2  \frac{\partial f_r}{\partial y}^2 \frac{\partial^2 f_r}{\partial x \partial y}^2\\
&\quad 
+12 v_s  \frac{\partial f_r}{\partial y}^2 \frac{\partial^2 f_r}{\partial x \partial y}^2-16 v_s f_r \frac{\partial^2 f_r}{\partial y^2} \frac{\partial^2 f_r}{\partial x \partial y}^2+12 v_s  \frac{\partial f_r}{\partial y}^2 \frac{\partial^2 f_r}{\partial t \partial z}^2+4 v_s  \frac{\partial f_r}{\partial y}^2 \frac{\partial^2 f_r}{\partial t \partial y}^2+\\
&\quad 
16 v_s^3 f_r  \frac{\partial f_r}{\partial y}^3  \frac{\partial f_r}{\partial x} \frac{\partial^2 f_r}{\partial x \partial y}-32 v_s  \frac{\partial f_r}{\partial y} \frac{\partial^2 f_r}{\partial y^2}  \frac{\partial f_r}{\partial x} \frac{\partial^2 f_r}{\partial x \partial y}-12 v_s^3 f_r  \frac{\partial f_r}{\partial y}^4 \frac{\partial^2 f_r}{\partial x^2}+\\
&\quad 
16 v_s f_r \frac{\partial^2 f_r}{\partial y^2}^2 \frac{\partial^2 f_r}{\partial x^2}+24 v_s^2 f_r  \frac{\partial f_r}{\partial y}^2 \frac{\partial^2 f_r}{\partial x \partial z} \frac{\partial^2 f_r}{\partial t \partial z}-16 \frac{\partial^2 f_r}{\partial y^2} \frac{\partial^2 f_r}{\partial x \partial z} \frac{\partial^2 f_r}{\partial t \partial z}+\\
&\quad 
16 v_s^2  \frac{\partial f_r}{\partial y}^3  \frac{\partial f_r}{\partial x} \frac{\partial^2 f_r}{\partial t \partial y}+8 v_s^2 f_r  \frac{\partial f_r}{\partial y}^2 \frac{\partial^2 f_r}{\partial x \partial y} \frac{\partial^2 f_r}{\partial t \partial y}-16 \frac{\partial^2 f_r}{\partial y^2} \frac{\partial^2 f_r}{\partial x \partial y} \frac{\partial^2 f_r}{\partial t \partial y}+\\
&\quad 
4 (4 \frac{\partial^2 f_r}{\partial y^2}^2-3 v_s^2  \frac{\partial f_r}{\partial y}^4) \frac{\partial^2 f_r}{\partial t \partial x}+4 \frac{\partial^2 f_r}{\partial z^2}^2 (4 (v_s  \frac{\partial f_r}{\partial x}^2+v_s f_r \frac{\partial^2 f_r}{\partial x^2}+\frac{\partial^2 f_r}{\partial t \partial x})-v_s  \frac{\partial f_r}{\partial y}^2)\\
&\quad 
-8 \frac{\partial^2 f_r}{\partial z^2} (v_s \frac{\partial^2 f_r}{\partial y^2}  \frac{\partial f_r}{\partial y}^2+4 v_s  \frac{\partial f_r}{\partial x} \frac{\partial^2 f_r}{\partial x \partial y}  \frac{\partial f_r}{\partial y}+2 (v_s f_r (\frac{\partial^2 f_r}{\partial x \partial z}^2+\frac{\partial^2 f_r}{\partial x \partial y}^2)+\\
&\quad 
\frac{\partial^2 f_r}{\partial x \partial z} \frac{\partial^2 f_r}{\partial t \partial z}+\frac{\partial^2 f_r}{\partial x \partial y} \frac{\partial^2 f_r}{\partial t \partial y}-2 \frac{\partial^2 f_r}{\partial y^2} (v_s  \frac{\partial f_r}{\partial x}^2+v_s f_r \frac{\partial^2 f_r}{\partial x^2}+\frac{\partial^2 f_r}{\partial t \partial x})))\biggl)
\end{aligned}
\label{r2long}
\end{equation}

\vspace{1cm}

%% please use the heading ``Extended Data''.    %%
%%=============================================%%

%%=============================================================%%
%% Sample for another appendix section			        %%
%%=============================================================%%


\end{appendices}

%%===========================================================================================%%
%% If you are submitting to one of the Nature Portfolio journals, using the eJP submission    %%
%% system, please include the references within the manuscript file itself. You may do this   %%
%% file, and delete the associated \verb+\bibliography+ commands.                             %%
%%===========================================================================================%%

\bibliography{sn-bibliography}% common bib file

\end{document}